\documentstyle[12pt,aaspp]{article}
\newcommand{\rb}[1]{\raisebox{1.5ex}[0pt]{#1}}
\newcommand{\Ref}{\hangindent=20pt \hangafter=1 \noindent}
\newcommand{\StartRef}{\hyphenpenalty=10000 \raggedright}
\newcommand{\beq}{\begin{equation}}
\newcommand{\eeq}{\end{equation}}

\newcommand{\NarrowMargins}{
  \setlength{\oddsidemargin}{+0.3in}
  \setlength{\evensidemargin}{-0.0in}
  \setlength{\textwidth}{6.2in}
  \setlength{\topmargin}{-0.75in}
  \setlength{\textheight}{9.25in}   }
\catcode`\@=11 
\def\lsim{\mathrel{\mathpalette\@versim<}}
\def\gsim{\mathrel{\mathpalette\@versim>}}
\def\@versim#1#2{\vcenter{\offinterlineskip
        \ialign{$\m@th#1\hfil##\hfil$\crcr#2\crcr\sim\crcr } }}
\catcode`\@=12 
\NarrowMargins
\begin{document}
\title{Scaling Laws for Advection Dominated Flows: Applications to
Low Luminosity Galactic Nuclei.}
\author{Rohan Mahadevan}
\affil{Harvard-Smithsonian Center for Astrophysics,
60 Garden St., Cambridge, MA 02138.}
 
\begin{abstract}
We present analytical scaling laws for self--similar 
advection dominated flows. 
The spectra 
from these systems range from 10$^{8}$ -- 10$^{20}$ Hz,  
and are determined by considering cooling 
of electrons through synchrotron, 
bremsstrahlung, and Compton processes. 
We show that the spectra can be quite accurately reproduced without
detailed numerical calculations, and that there is a strong 
testable correlation between the radio and X-ray fluxes from these
systems.  We describe how different 
regions of the spectrum scale with the mass of the accreting
black hole, $M$,  the accretion rate of the gas, $\dot{M}$, 
and the equilibrium temperature of the electrons, $T_e$. We show
that the universal radio spectral index of 1/3 observed in
most elliptical galaxies (Slee et al. 1994) is a natural 
consequence of self--absorbed synchrotron radiation from these
flows. We also 
give expressions for the total luminosity of these 
flows, and the critical accretion rate,  
$\dot{M}_{\rm crit}$, above which the advection solutions 
cease to exist.  We find that  for most cases of interest
 the equilibrium 
electron temperature is fairly insensitive to $M$, $\dot{M}$,  and
parameters in the model.  We apply these results to 
low luminosity black holes in galactic nuclei.  We show 
that the problem posed by Fabian \& Canizares (1988) 
of whether bright elliptical galaxies host
dead quasars is resolved, 
as pointed out recently by Fabian \& Rees (1995),
by considering advection--dominated flows. 
\end{abstract}
\section{Introduction.}
The observational proof for the existence of black holes is  
one of the outstanding  problems in astrophysics today.  It is 
generally believed that black holes exist in binary star systems, 
at the centers of most normal galaxies, and are the central 
engines that power distant quasars.  Attempts to prove the
existence of these singularities are confined to inferring their
presence by observing how they affect their environment.  
Measuring the kinematics of stellar systems and gas 
orbiting near the cores of galaxies (eg. van der Marel 1995a, b), 
using  time variability arguments of the X--ray
fluxes from quasars (eg. Wandel \& Mushotzky 1986), or measuring
the mass function in X-ray binaries (eg. Haswell et al. 1993)
 are ways of 
inferring the existence of a massive object (black hole) that  is 
confined to a small  volume.

Another way of inferring the presence of  a black hole is to consider
the emission spectrum produced by an accretion 
disk as the surrounding gas accretes onto the central object.  When 
considering black hole systems,  the standard
theory of accretion disks has serious difficulties in 
explaining the entire spectrum of these systems.  The primary 
problem in the standard  thin disk models is that the accreting gas
is optically thick, and radiates locally as a modified black 
body spectrum (see Frank et al. 1992).  This simple spectrum 
clearly falls short of explaining the entire emission from the
radio to hard X-rays of  these systems.  Models have been 
proposed which explain the emission spectrum at certain frequencies
(eg. Duschl \& Lesch  1994), but these fail to explain
the emission in other regions of the spectrum.

A possibility of explaining the entire spectrum of these systems
has recently emerged with the consideration of advection--dominated
accretion (Rees et al. 1982; Abramowicz et al. 1988; 
Narayan \& Yi 1994, 1995a,b; Abramowicz et al. 1995). 
Unlike standard accretion disk theory, one class of 
advection--dominated accretion considers accretion flows that are  
optically thin  and have low radiative efficiency.
These flows have a two--temperature structure (Shapiro, Lightman, \&
Eardley 1976) and hence do not require all the viscously 
dissipated energy to be radiated locally, but instead allow
a large fraction of the generated energy to be advected inwards,
with  the flow, to be ultimately deposited into the black hole.  The
total luminosity from these disks is therefore much lower,
for a fixed accretion rate, than the luminosity from a thin
accretion disk.  It is  however also possible to have
a disk structure where there is an outer thin disk, 
which becomes advection--dominated as the flow approaches the 
black hole.  In this case the outer disk gives the standard modified
black body spectrum (Frank et al. 1992) which produces standard
thin disk luminosities (eg. Narayan 1996; Narayan,  Mc Clintock \& 
Yi 1996, Lasota et al. 1996).  For the present discussion we neglect
the outer disk component since standard thin disks are well 
understood, and we are mainly interested in the advection--dominated
flow. 

The optically thin accretion flows in advection dominated
systems naturally require electrons in the gas to cool 
via synchrotron, bremsstrahlung and inverse Compton processes. 
These processes are responsible for
producing the entire spectrum, in these systems,   
from the radio to hard X-rays, in a natural way.  
A unique feature in considering advection flows to describe 
accreting black hole systems, is that 
they {\em require} the existence of an event horizon 
(Narayan, Yi, \& Mahadevan 1996; Narayan, Mc Clintock, \& Yi 1996),
since a hard surface (eg. a neutron star) would re--radiate all the 
advected energy, thereby producing an equivalent total luminosity as
predicted by a thin accretion disk. Successful application of 
these models to black hole systems might therefore prove
the existence of an even horizon 
(Narayan, Yi, \& Mahadevan 1995; Narayan, Mc Clintock, \& Yi 1996).

Detailed numerical calculations which consider the individual 
cooling and heating processes in the flow have been performed by 
Narayan \& Yi (1995b), and  the resulting spectra have been 
successfully applied to a number of putative black hole systems
(eg. Narayan et al. 1995, Lasota et al. 1996, Narayan, Mc Clintock,
\& Yi 1996).
Narayan \& Yi (1995b) have numerically 
obtained a number of 
interesting properties of these advection flows.
From the detailed calculations, however, it is difficult to 
deduce how different regions of a spectrum scale as quantities 
such as the mass of the central object and accretion rate are 
varied.  

The present paper develops analytical expressions to describe
the general
properties of advection dominated flows.
We deduce scaling laws which give physical 
insight to the detail processes involved, and show how these simple
laws give rise to quite an accurate description of these flows.
In \S \ref{pequationstex} we review the self--similar flow 
equations for advection dominated disks.  \S \ref{phtex}  describes
the heating and cooling processes, and \S \ref{psynchtex} shows
how the entire spectrum from these systems can be understood by 
simple scaling laws.  \S \ref{ptemperaturetex} addresses the general 
properties of the flow.  In \S \ref{fabiandis} we follow 
Fabian \& Rees (1995) and apply the results
to resolve  the  long standing problem  posed by
Fabian \& Canizares (1988) of whether elliptical galaxies host 
dead quasars.  Finally, in \S \ref{conclude.tex}, we discuss future
applications of these models and conclude.
\section{Self-Similar Flow Equations.}
\label{pequationstex}
In this section we review some of the basic assumptions and 
equations of the
self--similar advection dominated models developed by 
Narayan \& Yi (1995b).   Narayan \& Yi (1995b) 
present self--similar equations which describe local 
properties of the accreting gas as a function of the mass, $M$,  the 
accretion rate, $\dot{M}$,   the radius, $R$, the viscosity parameter,
$\alpha$, the ratio of gas pressure to total  pressure, $\beta$, and
the fraction of viscously dissipated energy that is  advected, $f$.

The accreting gas in an advection--dominated flow is a two temperature
optically thin plasma.  The ions are at their virial temperature
and the electrons are significantly cooler.   
The  total pressure, $p$, in these flows is the sum of 
gas ($p_g$) and magnetic ($p_m$) pressure.
The gas is roughly in equipartition with an isotropically
tangled magnetic field, $B$, which contributes a factor $1-\beta$ to the
total pressure,
\beq
p_m \equiv (1-\beta) \rho c_s^2 = {B^2 \over 24 \, \pi}.
\eeq
This equation differs from Narayan \& Yi (1995) by a factor of
1/3 to account for the pressure due to a three dimensional tangled
magnetic field. $\rho $ and $c_s$ are the mass density and speed of 
sound.
 
The self--similar equations are written in terms of scaled quantities:
the mass is scaled in solar mass units
\beq
M = m \, M_{\odot},
\eeq
the radius in Schwarzschild radii
\beq
R = r R_{\rm Schw}, \mbox{\hspace{1cm}} 
R_{\rm Schw} = { 2 G M \over c^2} = 2.95\times 10^5 \, m \ \ \mbox{cm},
\eeq
and the accretion rate in  Eddington units
$$\dot{M} = \dot{m} \, \dot{M}_{\rm Edd},$$
\beq
 \dot{M}_{\rm Edd} = { L_{\rm Edd} \over \eta_{\rm eff} c^2} = 
1.39\times 10^{18} m \ \ \mbox{g s$^{-1}$},
\eeq
where $\eta_{\rm eff}=0.1$ is the standard efficiency in converting 
matter to energy (Frank et al. 1992). 

Since these flows are essentially spherical in geometry 
(Narayan \& Yi 1995b), the vertical scale height of the disk is set equal
to the radius in the equations that follow.  
With this approximation and the scalings above, the self--similar 
equations for the accretion flow 
which are relevant for the present discussion are (Narayan \& Yi 1995b):
$$\rho = 6.00\times 10^{-5} \ \alpha^{-1} c_1^{-1} \,
                m^{-1} \, \dot{m} \, r^{-3/2} \ \ \mbox{g cm$^{-3}$},$$
$$B = s_1 \, m^{-1/2} \, \dot{m}^{1/2} \, r^{-5/4} \ \ \mbox{G},$$ 
$$n_e =b_1 \,  m^{-1} \, \dot{m} \, r^{-3/2}\ \ \mbox{cm$^{-3}$}, $$
$$ s_1 = 1.42\times10^9 \, \alpha^{-1/2} (1-\beta)^{1/2} c_1^{-1/2}
        c_3^{1/2}, $$
\beq
 b_1 =  3.16\times 10^{19} \, \alpha^{-1} \,
        c_1^{-1}. \label{alleq}
\eeq
These are the equations that differ from Narayan \& Yi (1995b) 
since we have assumed spherical accretion. 
$n_e$ is the
numberdensity of electrons, and $c_1$, $c_3$   are constants as 
defined in Narayan \& Yi (1995b). 
\footnote{ In the definition of 
$c_1, \ c_3$ as given in Narayan \& Yi (1995b), the ratio of 
specific heats of the gas is different from the present paper.  
We use (Esin 1996)
$$\gamma = {8 - 3 \beta \over 6 - 3\beta}.$$ }
For all cases of interest, $c_1 \simeq 0.5$ and $c_3 \simeq 0.3$.
\section{Energy Balance and Heating of a Two Temperature Plasma.}
\label{phtex}
The accreting gas in the advection flows are heated locally by
viscous forces.  In the analysis of Narayan \& Yi (1995b), 
the viscously dissipated energy $q^+$
 is mainly transferred to the ions in the gas.
A fraction $f$ of this energy is carried
 inwards by the accreting gas, while the remaining fraction $1-f$ is
transferred from the ions to the electrons to be radiated via 
synchrotron, inverse Compton and bremsstrahlung emission.  
There are therefore two energy equations that need to be satisfied.  
In the present analysis  we account for the possibility of 
viscously heating the electrons by a fraction $\delta$. 
Since the  
heat generated by the viscous forces is transferred  mainly to 
those particles with more inertial mass, we would expect 
that the fraction $\delta$ of viscous energy transferred to the
electrons is in the ratio of the electron to ion mass $\sim m_e/m_i
\sim 1/2000$.  
The energy balance for the ions therefore satisfies
\begin{eqnarray}
q^+  &=& f\, q^{+}  + q^{ie} + \delta \, q^+, \nonumber \\
&\equiv& q^{\rm adv} + q^{\rm e+}, \ \ \mbox{ergs s$^{-1}$ cm$^{-3}$},
\end{eqnarray}
with $$ q^{\rm adv}  \equiv  f \, q^+,$$  $$ q^{\rm e+} \equiv
q^{ie} + \delta \, q^+.$$
Here,  $q^+ $ and $q^{ie}$ are the rate of heating per unit volume and
rate of transfer of energy from the ions to the electrons per unit
volume respectively, $q^{\rm adv}$ is the advected energy, and 
$q^{\rm e+}$ is total electron heating rate including 
viscous heating. 

The electrons satisfy the energy equation, $q^{\rm e+} =  q^-$, 
where $q^-$ is the sum of all the local cooling processes (synchrotron,
bremsstrahlung, and inverse Compton). 
Setting $\delta = 0$
in these equations gives the energy equations of Narayan \& Yi (1995b).
 For a given $m$, $\dot{m}$, $r$, $\alpha$, and $\beta$, the electron
and ion energy equations are solved to determine the
electron and ion temperatures of the plasma, and to determine 
the fraction $f$ of advected energy.  Narayan \& Yi (1995b) use 
detailed numerical methods to solve these equations at each 
radius $r$, in order to  
determine the local properties of the flow and the spectrum that is
produced.
We obtain similar results with less effort analytically.

For the present analysis, the quantities of interest 
are the volume integrated quantities,
$Q^+, \ Q^{\rm e+}, \ Q^-$, which are obtained by integrating $q^+, \ 
q^{\rm e+}, \ q^-$ throughout the volume of the advection region. 
Using scaled quantities, and the approximation $H = R$, the volume
integrated quantities are defined by
\begin{eqnarray}
Q^X &=& \int_{R_{min}}^{R_{max}} 4 \pi \, R^2 \, q^X \, dR, 
\nonumber \\
&=& 3.23 \times 10^{17} \, m^3 \, \int_{r_{\rm min}}^{r_{\rm max}}
q^X \, r^2 \, dr \ \ \mbox{ergs s$^{-1}$}, \label{totaldef}
\end{eqnarray}
where $X$ denotes any quantity of interest.  The lower 
limit is taken to be $r_{\rm min} = 3$ since the self--similar
solutions break down for $r \lsim $few (Mastsumoto et al. 1985, 
Narayan 1996).  This choice of $r_{\rm min}$ is also 
in accordance with previous calculations (eg. Narayan et al. 1995, 
 Lasota et al. 1996), and we find that
this reproduces the detailed spectra quite
well.  To 
determine the upper limit, we use some of the properties of the
flow developed by Narayan \& Yi (1995b).  
Narayan \& Yi (1995b) have shown that
for $r \gsim 10^3$, the flow becomes a cool $\sim 10^{8.5}$ K
one temperature plasma, and not much radiation is produced, 
 while for $r < 10^3$ the electron
temperature is fairly constant while the ion temperature increases
as $1/r$.   
Since most of the radiation from these flows originate
at $r<10^3$, where the $T_e \gsim 10^9$, 
and the present discussion is interested in the 
radiation produced from such a flow, we set $r_{\rm max} = 10^{3}$.   
In the discussion that follows, we assume that the 
electron temperature is constant for $r < 10^3$, as suggested by
the detailed calculations of Narayan \& Yi (1995b).
The energy balance equations take the form
$$Q^+ = Q^{\rm adv} + Q^{\rm e+},$$
$$Q^{\rm e+} = Q^{ie} + \delta \, Q^{+},$$
$$Q^{\rm e+} = Q^{-}, $$
\beq
Q^{-} = P_{\rm synch} + P_{\rm Compton} + P_{\rm brems},  
\label{energyQeq}
\eeq
where $P_{\rm synch}, \ P_{\rm Compton}, \  P_{\rm brems}$ are the
total cooling rates for the individual processes. 
The energy equations for the ions and electrons are solved
self--consistently to determine the fraction of the advected 
energy $f$, and the electron temperature $T_e$. To do this, we 
first give analytic equations for
the heating terms $Q^+, \ Q^{ie}$, and in the next section determine
the cooling terms $P_{\rm synch}, \ P_{\rm Compton}, \  P_{\rm brems}$,
and the spectra they produce.

\subsection{Heating Processes: Ion Heating}
The ions are heated by viscous forces.   The total heating rate, 
$Q^+$, is obtained by using eq.(\ref{totaldef}) and  integrating $q^+$, 
as defined in Narayan \& Yi (1995b),
throughout the advection region.  This gives
\beq
Q^+ = 9.39\times 10^{38} \ {1 - \beta \over f} \, 
        c_3 \, m \, \dot{m} \, r_{\rm min}^{-1} 
\ \ \mbox{ergs s$^{-1}$}, \label{qplus}
\eeq
where we have set $r_{\rm max} \gg r_{\rm min}$.
For low values of $\alpha$, $c_3$ is independent of $\alpha$, and 
eq. (\ref{qplus}) shows that for fixed  
$m$ and $\dot{m}$, 
the heating rate depends only on 
the fraction of gas to magnetic pressure. 

\subsection{Heating Processes: Electron Heating.} 
The electrons are heated by two processes: 
by viscous heating $\delta \, Q^+$, where an expression for 
$Q^+$ has been derived above, 
and by a transfer of energy from the ions to 
electrons via Coulomb interactions.  The heating rate per unit 
volume due to 
Coulomb interactions is given by 
Stepney \& Guilbert (1983), Narayan \& Yi (1995b), and can be approximated
to (see Appendix \ref{appendixqie})
\beq
q^{ie} \simeq 5.61\times 10^{-32} \, (T_i - T_e) \, 
b_1^2 \, m^{-2} \, \dot{m}^{2} \,
        r^{-3} \,  r^{-1} \, g(\theta_e)
\ \ \mbox{ergs cm$^{-3}$ s$^{-1}$,}
\eeq 
where we have substituted for $n_e$, $\theta_e = k T_e/m_e c^2$, and
\beq
g(\theta_e) \equiv {1 \over
                K_2(1/\theta_e)} \, \left(2 + 2\theta_e + { 1 \over
                \theta_e} \right) \, e^{-1/\theta_e},
\eeq
which is tabulated for various values of temperature in
 Table \ref{k2table}.  From Narayan \& Yi (1995b), the ion temperature
can be approximated to 
\begin{eqnarray}
T_i &=& 6.66\times10^{12} \beta c_3 r^{-1} - 1.08 T_e, \nonumber \\
&\simeq& h \, r^{-1}, \label{ti}
\end{eqnarray}
where $$ h = 6.66 \times 10^{12} \beta c_3. $$ The 
the second term  in eq.(\ref{ti}) has been neglected compared 
with the first 
since the electron temperatures are considerably lower that the 
ion temperatures for $r\lsim 10^3$.

The total ion--electron heating rate for the electrons is (cf. eq. 
\ref{totaldef})
\beq
 Q^{ie} \simeq 1.2\times 10^{38} \, g(\theta_e) \,
\alpha^{-2} \, c_1^{-2} \, c_3 \, 
 \beta \, m \, \dot{m}^2 \, r_{\rm min}^{-1}
\ \mbox{ergs s$^{-1}$},  \label{qiemcrit}
\eeq
where we have substituted for $b_1$, $h$, and  assumed 
$r_{\rm max} \gg r_{\rm min}$.
Combining the equations above, the total heating of the electrons is 
given by
\begin{eqnarray}
Q^{\rm e+} &=& Q^{ie} + \delta \, Q^{+}, \nonumber \\
&\simeq& 1.2\times 10^{38} \, g(\theta_e) \,
\alpha^{-2} \, c_1^{-2} \, c_3 \, \beta \, m \, \dot{m}^2 \, r_{\rm min}^{-1}
\nonumber \\
&+& \delta \, 
9.39\times 10^{38} \ \epsilon^{\prime} \,
        c_3 \, m \, \dot{m} \, r_{min}^{-1}. \label{qe+total}
\end{eqnarray}
The major source for electron heating depends on the value of 
$\dot{m}$; for high $\dot{m}$, $Q^{ie} \gg  \delta \, Q^+$, whereas 
for low $\dot{m}$, $\delta \, Q^{+} \gg Q^{ie}$. By setting
$Q^{ie} = \delta \, Q^+$, we can determine the transition 
accretion rate:
\beq
\dot{m} \sim 8.8 \times 10^{-5} \,  
\left( {\alpha \over 0.3} \right)^2
\left( {\delta \over 2000^{-1}} \right)
\left( {1 - \beta \over 0.5} \right)
\left( {\beta \over 0.5} \right)^{-1}
\left( {c_1 \over 0.5} \right)^2
\left( {f \over 1.0} \right)^{-1} \, g(\theta_e)^{-1}.\label{qe+comp}
\eeq
\section{Energy Balance: Cooling Processes and the 
Components of the Spectrum}
\label{psynchtex}
In order to balance the viscous and Coulomb heating, the electrons
cool through three distinct processes: synchrotron, bremsstrahlung,
and inverse Compton emission.
The emission  in
different regions of the spectrum is determined by these
individual cooling processes.  Synchrotron radiation 
is responsible for the radio to sub-millimeter emission, while a 
combination of bremsstrahlung emission and  inverse 
Compton scattering of 
synchrotron photons is responsible for the sub-millimeter to 
X-ray emission.
This is one of the successes of the advection-dominated  
models: explaining, using few free parameters, 
the entire spectra of accreting systems.
A natural question to ask is how does the
amount of emission and shape of the final 
spectrum depend on variables like
$\alpha$, $\beta$, $m$, $\dot{m}$, and  $T_e$ ?  
Previous papers (eg. Narayan \& Yi 1995b; 
Narayan, McClintock, \& Yi 1996)
have used detailed numerical calculations to evaluate the
spectrum produced. The analysis presented here give less detailed
spectra, but is much faster in determining the general 
characteristics, and the individual components of the spectra 
produced.

In the analysis that follows, the spectrum is 
divided into three components.  
The cyclo-synchrotron component, 
and the bremsstrahlung 
and  the inverse Compton component. 
Fig. 1 shows representative plots of the spectrum for 
a fixed mass  $m= 5\times 10^9$, and for different accretion
rates $\dot{m} = (3, 6, 12, 24)  \times 10^{-4}$, with 
$\alpha = 0.3$, and $\beta = 0.5$. For one curve, the 
individual components
of the spectrum have been labeled as S for synchrotron, 
B for bremsstrahlung,
and C for comptonization.
In the sub--sections below each of these components are describe
with the appropriate analytic approximations.  

\subsection{Cyclo-Synchrotron Emission and the Radio-Sub-mm Spectrum.}
The radio to sub-mm spectrum is defined by three quantities: 
1) the luminosity of the radio spectrum, 
2) the maximum (peak) frequency beyond which the spectrum falls off 
exponentially, and
3) the slope of the radio spectrum.  
We treat each of these separately.

In the optically thin limit, the spectrum of cyclo--synchrotron 
radiation by an isotropic distribution of relativistic thermal 
electrons is given by (Mahadevan et al. 1996, Narayan \& Yi 1995b)
\beq
\epsilon_{\rm synch} d\nu = 
4.43\times 10^{-30} {4 \pi \, n_e \, \nu \over K_2(1/\theta_e)}
\, M(x_M),
\eeq
where we use the extreme relativistic expression for $M(x_M)$ given
by 
\beq
M(x_M) = {4.0505 \over x_M^{1/6}} \, 
\left( 1 + { 0.40 \over x_M^{1/4}} + {0.5316 \over x_M^{1/2}} 
\right) \, \exp\left( -1.8899x_M^{1/3}\right) ,
\eeq
and 
\beq
x_M \equiv {2 \nu \over 3 \nu_b \theta_e^2}, \mbox{\hspace{1cm}}
\nu_b \equiv {e B \over 2 \pi \, m_e \,  c}.
\label{def1}
\eeq
The cyclo--synchrotron photons in these plasmas are self-absorbed, and
give a black body spectrum, up to 
a critical frequency $\nu_c$.  
The frequency at which this occurs, at each radius $r$,
is determined by evaluating the total cyclo-synchrotron emission over
a volume of radius $r$, and equating it to the Raleigh-Jeans black
body emission from the surface of this sphere.   This gives the
condition
\beq
4.43\times 10^{-30} {4 \pi \, n_e \, \nu_c \over K_2(1/\theta_e)}
\, M(x_M) \, {4 \pi \over 3} R^3 = \pi 2 {\nu_c^2 \over c^2}
\, k T_e \, 4\pi \, R^2
\eeq
which can be rewritten in terms of $x_M$ as
\beq
\exp\left(1.8899 \, x_M^{1/3}\right) = 2.49 \times 10^{-10} \
        {4 \pi \, n_e \, R \over B} \, {1 \over \theta_e^3 \,
                K_2(1/\theta_e) } \, \left( {1 \over x_M^{7/6}}
                        + {0.40 \over x_M^{17/12}} + {0.5316 \over
                                x_M^{5/3}} \right). \label{xmeq}
\eeq
$x_M$ is determined in Appendix \ref{xmappendix}.
Given $x_M$, the
cutoff frequency at each radius is determined by eqs.(\ref{def1}) to be 
\begin{eqnarray}
\nu_c &=& {3 \over 2 } \, \theta_e^2 \, \nu_b \, x_M, \nonumber \\
&=& s_1 \, s_2 \, m^{-1/2} \, \dot{m}^{1/2} \, T_e^2 \, r^{-5/4},
\ \ \mbox{Hz},
\label{cutoff}
\end{eqnarray}
where $s_1$ is given in eqs. (\ref{alleq}) and 
\beq 
s_2 \equiv 1.19\times 10^{-13} \, x_M,
\eeq
At this frequency the radiation becomes optically thin and
 the luminosity is given
by the Raleigh-Jeans part of the black body spectrum
\begin{eqnarray}
L_{\nu_c} &=& \pi \, 2\, {\nu_c^2 \over c^2} \,
k \, T_e \, 4\pi \, R^2, \nonumber \\
&=& s_3 \, T_e \, \nu_c^2 \, m^2 \, r^2 \ \ \mbox{ergs s$^{-1}$ Hz$^{-1
}$}, \ \ \ \ s_3 = 1.05\times10^{-24}.
\label{lnu}
\end{eqnarray}
This determines the luminosity at each point along the radio spectrum. 

For a fixed $T_e$, eq.(\ref{cutoff}) shows how the cutoff frequency 
varies with  $r$. Emission observed at higher frequencies originates
at smaller radii, closer to the central black hole. The peak 
frequency, and the power at that frequency  
are  determined by setting $r = r_{\rm min}$ in
eqs.(\ref{cutoff}),(\ref{lnu}),
$$\nu_p = s_1 \, s_2 \, m^{-1/2} \, \dot{m}^{1/2} \, T_e^2 \,
                r_{\rm min}^{-5/4} \ \ \mbox{Hz}, $$
\beq
\nu_p L_{\nu_p} =   s_1^3 \, s_2^3 \, s_3 \, r_{\rm min}^{-7/4} \,
                m^{1/2} \, \dot{m}^{3/2} \, T_e^7
                \ \ \mbox{ergs s$^{-1}$},
\eeq
which shows that the  luminosity at the peak frequency 
is very sensitive to the electron temperature.

The slope of the radio spectrum is a direct consequence of 
$x_M$ and $T_e$ being essentially constant, and the 
the Raleigh--Jeans part of the black body spectrum.
Since $B\propto r^{-5/4}$, eq.(\ref{cutoff}) shows that 
$r \propto \nu_c^{-4/5}$.  From eq.(\ref{lnu}),  $L_{\nu} \propto \nu_c^2 \,
r^2 \propto \nu_c^{2/5}$.  
The complete spectrum is obtained by 
rewriting eq.(\ref{cutoff}) in terms of $r$
and substituting in eq.(\ref{lnu}) to give
\beq
L_{\nu}  \simeq  s_3 \, (s_1 s_2)^{8/5} \ m^{6/5} \dot{m}^{4/5} 
\, T_e^{21/5} \, \nu^{2/5} \ \ \mbox{ergs s$^{-1}$ Hz$^{-1}$}.
\label{rvnt2}
\eeq
This produces a spectrum with slope of $2/5$, which is similar to 
the slope of 1/3 produced by optically thin synchrotron emission
(the dependence of $x_M$ on $r$ is not taken into account
here, numerically, $x_M \sim r^{1/15}$ which makes $L_{\nu}$ even 
closer to $\sim \nu^{1/3}$).  The advection--dominated models therefore
give a very natural explanation to the characteristic 1/3 radio spectral
indices observed when looking at putative black hole systems 
(Wrobel 1991; Slee et al. 1994; Narayan et al. 1995).
The 2/5 spectral slope extends from $\nu_p$ down
to $\nu_{\rm min}$ where $\nu_{\rm min}$ is the cutoff frequency 
given by setting $r = r_{\rm max} $ in eq.(\ref{cutoff}) (cf.
Fig. 1).  This is
is a direct consequence of the advection flows having a constant 
electron temperature  for $r \lsim 10^3$. Beyond this radius the
advection flows become a one temperature plasma and
$T_e \propto r^{-1}$, which gives a steeper radio slope of 
$ 22/13$, as long as $T_e \gsim
 10^8$K (below this temperature there is no  synchrotron radiation).

 We assume that $\nu_{\rm min} \ll \nu_p$ and obtain 
the total power from  
\begin{eqnarray}
P_{\rm synch} &=& \int_{0}^{\nu_p} \, L_{\nu} \, d\nu \simeq \, 0.71 \, \nu_p \, L_{\nu_p}, \nonumber \\
&\simeq& 5.3\times 10^{35} \,
\left( {x_M \over 1000} \right)^3
\left( {\alpha \over 0.3} \right)^{-3/2}
\left( {1-\beta \over 0.5} \right)^{3/2}
\left( {c_1 \over 0.5} \right)^{-3/2}
\nonumber \\
&\times&
\left( {c_3 \over 0.3} \right)^{3/2}
\left( {r_{\rm min} \over 3} \right)^{-7/4}
\left( {T_e \over 10^9} \right)^{7}
\, m^{1/2} \, \dot{m}^{3/2}
\ \ \mbox{ergs s$^{-1}$}.
\label{psynch}
\end{eqnarray}
In this simple description, the synchrotron spectrum is assumed to 
terminate at $\nu_p$  
(cf. Fig. 1) 
and does not reproduce the exponential 
decay 
in the emission that is expected from thermal plasmas 
(Mahadevan et al. 1996).  This is because, in this simple 
discussion,  we have
assumed that all the photons to be comptonized occur at the
peak frequency (see below), and so the comptonized spectrum begins
at $\nu_p$.
\subsection{Bremsstrahlung Emission: The Sub-mm to 
Hard X-ray Spectrum.}
Bremsstrahlung emission is characterized by a constant luminosity 
$L_{\nu}$, up to a frequency $h \nu = k \, T_e$ where the spectrum
turns and falls off exponentially (cf. Fig. 1).  
The total emission due to bremsstrahlung 
radiation is given by  eq. (\ref{totaldef}) with $q^{X} = q_{\rm brems}$, 
where
$q_{\rm brems}$ is the bremsstrahlung emission per unit volume due
to both electron-electron and electron-ion interactions.  
The bremsstrahlung emission
per unit volume is given by (Stepney \& Guilbert 1983; Narayan \& Yi 1995b), 
\begin{eqnarray}
q_{\rm brems} &=& q_{ei} + q_{ee}, 
\nonumber \\
&\simeq& 1.48 \times 10^{-22} \, n_e^2 \, F(\theta_e), 
 \ \ \mbox{ergs cm$^{-3}$ s$^{-1}$},
\end{eqnarray}
which represent the contributions from electron--electron and
electron--ion interactions, and 
\beq
F(\theta_e) =
\left\{ \begin{array}{l@{\quad  \quad}l}
        4 \left( {\displaystyle{2 \theta_e \over  \pi^3}}
\right)^{1/2} \! ( 1 +
                1.781 \, \theta_e^{1.34}) +
                 1.73 \, \theta_e^{3/2}( 1 + 1.1 \, \theta_e +
                        \theta_e^2 - 1.25 \, \theta_e^{5/2}),
                & \theta_e < 1, \\
& \\
        \ \ \left(\displaystyle{{9 \theta_e \over 2 \pi}} \right) \
\left[\ln(1.123 \, \theta_e + 0.48) + 1.5 \right]
        + 2.30 \,  \theta_e(\ln 1.123 \, \theta_e + 1.28), &
        \theta_e > 1, \end{array} \right.
\eeq
Using the expression for the number density, the 
total bremsstrahlung power is 
\beq
P_{\rm brems} 
= 4.78 \times10^{34} \, \alpha^{-2} \, c_1^{-2} \, 
\ln(r_{\rm max}/r_{\rm min}) \, 
  F(\theta_e) \, m \, \dot{m}^{2},
\label{Pb}
\eeq
and the spectrum due to bremsstrahlung emission is
\begin{eqnarray}
\lefteqn{L_{\rm brems}(\nu) \simeq 2.29\times 10^{24} \,
\, \alpha^{-2} \, c_1^{-2} \, 
\ln(r_{\rm max}/r_{\rm min})} \nonumber \\
&\times& F(\theta_e) \, T_e^{-1} \, e^{- h \, \nu/k\, T_e} 
\, m \, \dot{m}^{2}
 \ \ \mbox{ergs s$^{-1}$ Hz$^{-1}$},
\end{eqnarray}
which is shown in Fig. 1.  Comparing eq. (\ref{Pb}) with 
eq. (\ref{psynch}) shows that for most cases of interest, 
$P_{\rm brems} < P_{\rm synch}$, and can be neglected when considering
the total cooling rate of the flow.  However, when considering 
the entire emission spectrum, bremsstrahlung emission is important
since it contributes to the X-ray emission.
\subsection{Comptonization: The Sub-mm to Hard X-ray Spectrum.}
In this discussion, we neglect
the comptonization of bremsstrahlung emission, and only consider 
the comptonization of the soft cyclo--synchrotron 
photons. 
This is the other process responsible for the sub-mm to hard x-ray
spectrum. 
The spectrum is defined by three quantities: 1) the initial
frequency of the photons that are comptonized, 2) the maximum final 
frequency  of a comptonized photon, and 3) the slope,
$\alpha_c$, of the comptonized spectrum (cf. Fig. 1).  

The photons that are comptonized are the soft cyclo--synchrotron photons 
in the radio spectrum. The emission in the radio spectrum mainly occurs 
at the peak frequency, $\nu_p$,  and so we can make the approximation 
that all the synchrotron photons to be comptonized 
have an initial frequency of $\nu_p$.  
The maximum final frequency of a comptonized photon is 
$h \nu_f = 3 \, k \, T_e$, which is the average energy of a photon
for saturated comptonization in the Wien regime.

The optical depth to electron scattering, $\tau_{es}$, and
how much a photon is amplified in one scattering, are the two 
quantities that determine the slope of the Compton spectrum. 
Photons at different radii see different optical depths, with 
photons  at  small radii seeing large optical depths and those at  
large radii seeing small optical depths.
In this simple treatment we expect,
on the average, that all the photons would probably see half the total 
optical depth.
We therefore take the optical depth
to electron scattering to be half of that as 
given in Narayan \& Yi (1995b),
\begin{eqnarray}
\tau_{es} &=& 6.2 \, \alpha^{-1}  \, c_1^{-1} \, \dot{m} \, r^{-1/2}, 
\nonumber \\
&=& ( 23.87 \, \dot{m})
\left( {\alpha \over 0.3} \right)^{-1}
\left( {c_1 \over 0.5} \right)^{-1} 
\left( {r_{\rm min} \over 3}\right)^{-1/2}. \label{second}
\end{eqnarray}
We find that this choice of $\tau_{es}$ reproduces the more detailed 
comptonized spectrum quite well (Narayan, private communication).

In the standard treatment of comptonization (e.g. Rybicki \& Lightman 1979, 
Dermer, Liang, \& Canfield 1991), a photon with initial energy 
$\epsilon_i$ that undergoes $k$ scatterings, has final energy 
$\epsilon_f \simeq A^k \,  \epsilon_i$, where $A$ is the mean amplifaction
factor in one scattering which for thermal plasmas is
\beq
A = 1 + 4 \, \theta_e  + 16 \, \theta_e^2.
\eeq
For temperature ranges of interest, $2 < A < 50$. The luminosity of 
the emerging photons
at frequency $\nu_f$ has the power--law shape 
\beq
L_{\nu_f} \simeq L_{\nu_i} \, \left( {\nu_f \over \nu_i}\right)
^{-\alpha_c}, \label{lnucomp}
\eeq
where
\beq
\alpha_c \equiv { - \ln \tau_{es} \over \ln A}. \label{alphac}
\eeq
The total Compton power is 
\begin{eqnarray}
P_{\rm Compton} &=& \int_{\nu_p}^{3 \, k \, T_e/h} \, L_{\nu_f} \, d\nu_f,
\nonumber \\
&=&
{\nu_p \, L_{\nu_p}  \over 1 - \alpha_c}
\left[
\left( { 6.2 \times 10^{7} \, (T_e/10^9) \over  (\nu_p/10^{12})} 
\right)
^{1-\alpha_c}  - 1 \right]
\ \ \mbox{ergs s$^{-1}$}. \label{pcompton}
\end{eqnarray}
Eqs. (\ref{lnucomp}), (\ref{alphac}),  and (\ref{pcompton}) 
show how the Compton power depends on the optical depth and
temperature through the slope of the spectrum $\alpha_c$.
If $\alpha_c \gg 1$, comptonization can be neglected. 
If $\alpha_c \lsim 1$
then there is significant comptonization of the cyclo--synchrotron 
photons, and the cooling is dominated
by the inverse Compton losses.  The actual determination of 
$\alpha_c $ is done self--consistently and is discussed in 
\S \ref{ptemperaturetex}.  Although eq. (\ref{pcompton}) is 
used to determine the total Compton power in the subsequent 
sections (see \S \ref{ptemperaturetex}), it is instructive to 
analytically approximate eq. (\ref{pcompton}) for $\alpha_c \gg 1$,
and $\alpha_c <1$, to determine how the Compton power scales
in these regimes.  We consider these two cases below, and show
how the value of $\alpha_c$ determines whether the comptonization
of the soft--cyclosynchrotron photons dominates over 
bremsstrahlung emission in the 
sub-mm to X-ray region of the spectrum.

\subsubsection{$\alpha_c > 1$:  Low Compton Cooling.}
\label{LCC}
For $\alpha_c > 1$, the first term in the square brackets in 
eq. (\ref{pcompton}) can
be neglected which gives 
\beq
P_{\rm Compton}(\alpha_c>1)
\simeq {\nu_p \, L_{\nu_p} \over \alpha_c -1},
\label{alphag2comp}
\eeq
with $\alpha_c \gg 1$ corresponding to no comptonization.  Comparing
this with eq.(\ref{psynch}), the total Compton power is proportional 
to  the total synchrotron power. If $\alpha_c \gg 1 $, then the Compton
power is less than the synchrotron power, and can be neglected when 
determining the total cooling rate. When 
$ 1 < \alpha_c < 2 $, however, the Compton power is
 greater than the synchrotron power and  contributes comparably 
to the total cooling rate.

\subsubsection{$\alpha_c < 1$: Significant Compton Cooling.}
For $\alpha_c < 1$, the second term in square brackets in 
eq. (\ref{pcompton}) can be neglected which gives
\beq
P_{\rm Compton}(\alpha_c <1) \simeq
\left( { 6.2\times 10^{7} \, (T_e/10^9) \over \nu_p/10^{12}}\right)^{1-\alpha_c} {  \nu_p \, L_{\nu_p} \over  
1 - \alpha_c}.
\eeq
In this regime the Compton power dominates the total synchrotron power.
The Compton power when $\alpha_c = 1$ is obtained by taking
the limit as $\alpha_c \rightarrow 1$ of eq.(\ref{pcompton}). 

\subsubsection{Compton Luminosity.}
The luminosity in the sub-mm to X-rays due to comptonization is
\beq
L_{\rm Compton} \simeq \nu_p^{\alpha_c} \, L_{\rm synch}(\nu_p) \, 
\nu^{-\alpha_c} \ \ \mbox{ergs s$^{-1}$ Hz$^{-1}$},
\eeq
and is sensitive to whether 
$\alpha_c$ is less than, equal to, or greater than 1.   
For $\alpha_c \gg 1$, the bremsstrahlung luminosity in 
the sub-mm to X-rays is greater than the Compton luminosity.  When 
$\alpha_c <1$ comptonization dominates the sub-mm to X-ray spectrum, 
and when $1 < \alpha_c < 2$ both bremsstrahlung and comptonization
are dominant. Fig. 1 shows how an increase in the accretion rate 
increases the slope of the Compton spectrum. At low $\dot{m}$ the
bremsstrahlung emission  dominates the X-ray emission, when 
$\alpha_c > 1 $, whereas
for high $\dot{m}$, $\alpha_c <1$, and  
the comptonized spectrum dominates the X-ray emission. 
$\alpha_c$ therefore determines 
the dominant source of emission at these frequencies. Note that
in this simple description of comptonization, the spectrum begins
from $\nu = \nu_p$ (cf. Fig 1) and therefore does not reproduce the 
characteristic dip in the spectrum between radio and 
sub-millimeter wavelengths (eg. Narayan et al. 1995).  
A more detailed Compton  calculation would be needed for this.
\section{General Properties of the Flow.}
\label{ptemperaturetex}
In the following sections, we use the results obtained to determine
general properties of the advection--dominated
 flow.  We first determine the
temperature of the gas and $\alpha_c$, 
then  the total luminosity  from the flow, and
finally the critical accretion rate $\dot{m}_{\rm crit}$ above which
 the advection solution does not exist.
\subsection{Equilibrium Temperatures and $\alpha_c$.}
Since the electrons are responsible for cooling, the 
temperature in these flows is determined by the energy balance 
equation for the electrons.  The sum of the 
individual cooling processes
are equated to the total heating of the electrons, and this 
equation is solved self--consistently for the temperature.  
We first determine the equilibrium temperature and $\alpha_c$ through 
simple numerical methods, and then provide analytic 
approximations which determine them.
\subsubsection{Numerical Method.} 
\label{numericalmethod}
For a given $m$, $\dot{m}$, $\alpha$, and $\beta$, 
the total heating of the electrons is equated 
to the individual cooling processes, 
$Q^{e+} = P_{\rm synch} + P_{\rm brems} + P_{\rm Compton},$  
and the electron
temperature is varied  until this equality
is satisfied.  At each value of $T_e$, the slope of the 
comptonized spectrum is determined through eq.(\ref{alphac}).
Solving for the electron temperature therefore fixes the 
slope of the comptonized spectrum. Fig. 2 shows numerical plots of the equilibrium temperature as
a function of $\dot{m}$ for different values of the black hole
mass $m$.  Here, $\alpha = 0.3$, $\beta = 0.5$, and
$\delta = 1/2000$.   The corresponding values of $x_M$ at each
$\dot{m}$ is also shown.  
At high $\dot{m}$, the equilibrium temperatures
are independent of $m$ and are constant at a value
$T_e \simeq 2.0 \times 10^9$.  Further, at low $\dot{m}$,
$T_e$ increases with decreasing $\dot{m}$.  Note however
that if $\delta = 0$ then eq. (\ref{appendixte1}) shows that
the temperature decreases as $\dot{m}$ decreases.  This is because
synchrotron cooling is the dominant source of cooling, and is
much more efficient than the Coulomb heating at low $\dot{m}$.
 
Fig 3. shows the value of $1-\alpha_c$, the slope of the spectrum
on a $\nu \, L_{\nu}$ plot, as a function of $\dot{m}$,
for different values of the central mass $m$.
These correspond to the equilibrium conditions as shown in Fig. 2.
At low $\dot{m}$, $\alpha_c$ becomes constant which is expected since
both $\ln A$, $\ln \tau_{es}$ $\propto \ln \dot{m}$.  The value
of this constant depends on the mass of the central black hole. At
high accretion rates  $\alpha_c \sim 0.5$

\subsubsection{Analytic Determination.}
\label{tealphg1}
An analytic determination of the equilibrium electron 
temperature allows an 
understanding of how it scales with different 
quantities in these models.
 To simplify the 
equations that follow, note that eqs.(\ref{psynch}), 
(\ref{Pb}), and  (\ref{pcompton})  show that 
$P_{\rm brems} < P_{\rm synch}$, and that depending on the 
value  of $\alpha_c$, $P_{\rm synch}$ can be greater or 
less than $P_{\rm Compton}$. Further, since $P_{\rm brems}$ is very
insensitive to the electron temperature ($\propto F(\theta_e)$), as compared with $ (P_{\rm synch} +
P_{\rm Compton} \propto T_e^7)$, we find that for all ranges of 
$m, \  \dot{m}$, the contribution to the total cooling  
by bremmstrahlung emission, is negligible compared with  synchrotron and
Compton cooling, at the equilibrium temperatures.  We therefore neglect
bremmstrahlung cooling in the analysis that follows.

A rough estimate  
of $\alpha_c$, for a given $\dot{m}$,
 can be obtained from eqs.(\ref{second}) and  (\ref{alphac}).
Using the range of temperatures of interest
 ($10^9 \le T_e \le 2\times 10^{10}$) to determine the maximum and
minimum values of $\ln A$, and setting $\alpha_c = 1$, eqs. 
(\ref{alphac}), (\ref{second}) show that if 
$\dot{m} \lsim 10^{-4} \alpha$, then $\alpha_c \gsim 1$, and if 
$\dot{m} \ge 3\times 10^{-3} \alpha$ then $\alpha_c \le 1$.  
However for $ 10^{-4}\alpha \le \dot{m} \le 10^{-2}\alpha$ the value of 
$\alpha_c$ can be
either greater or less than  1, depending  on the temperature which 
has to be solved self-consistently.  We consider the two cases.

\noindent \underline{\bf$\alpha_c > 1$.} \\[0.25cm]
In this limit synchrotron and Compton emission are  
the dominant sources of
cooling, and depending on the value of $\alpha_c$ Compton 
cooling is comparable to or less than the total synchrotron cooling 
(cf. \S \ref{LCC}).
Bremsstrahlung cooling is neglected. 
The total cooling rate is given by
\begin{eqnarray}
Q^- &\simeq& \left( 0.71 + {1 \over \alpha_c -1} \right) \,
        \nu_p \, L_{\nu_p} \nonumber \\
&\simeq& A_c \, \nu_p \, L_{\nu_p},
\end{eqnarray}
where the first term is due to synchrotron cooling and the second is 
due to Compton cooling (for $\alpha_c \gg 1$ we only consider 
synchrotron cooling and $A_c = 0.71$). 
This has to be equal to the total heating $Q^{\rm e+}$.  However when
$\alpha_c > 1$ and $\dot{m} < 10^{-3} \, \alpha^2$, 
from eq.(\ref{qe+comp}) this
is when $Q^{ie} $ can be neglected compared with $\delta \, Q^+$ 
(see Appendix \ref{tedetappendix} for the case $\delta = 0$). 
Setting  $\delta \, Q^{+} = Q^{-}$,  and rearranging terms 
gives
\begin{eqnarray}
T_e &=&
{1.1 \times 10^{9} \over A_c^{1/7} }
\left( {\delta \over 2000^{-1}} \right)^{1/7}
\left( {x_M \over 300} \right)^{-3/7}
\left( {\alpha \over 0.3} \right)^{3/14}
\left( {1-\beta \over 0.5} \right)^{-1/14}
\nonumber \\
&\times&
\left( {c_1 \over 0.5} \right)^{3/14} 
\left( {c_3 \over 0.3} \right)^{-1/14}
\left( {r_{\rm min} \over 3} \right)^{3/28} 
 m^{1/14} \, \dot{m}^{-1/14} \ \ \mbox{K}, \label{alphg1}
\end{eqnarray}
where $A_c^{1/7}$ varies from 0.95 to 1.4, and we have scaled
$x_M$ appropriately for low $\dot{m}$ (cf. Fig. 2).
Fig. 2 shows the temperature increasing faster with $\dot{m}$ than
indicated above.  This is mainly due to the sensitivity of 
the temperature on $x_M$, which decreases since the synchrotron
emission decreases as $\dot{m}^2$.  However, comparing the four
panels in Fig. 2, shows that the temperature is fairly
insensitive to the mass of the central black hole. 

\noindent \underline{\bf$\alpha_c < 1$.} \\[0.25cm]
In this regime we find simple recursive formulae that can be
used to determine $T_e$ analytically.  
For $\alpha_c < 1$, both synchrotron and bremsstrahlung cooling is 
negligible, and the total cooling, $Q^-$ is given by 
eq.(\ref{pcompton}) 
\begin{eqnarray}
Q^{-} \simeq 
P_{\rm Compton} &=& {\nu_p \, L_{\nu_p}  \over 1 - \alpha_c}
\left[
\left( { 6.2 \times 10^{7} \, (T_e/10^9) \over  (\nu_p/10^{12})}
\right)
^{1-\alpha_c}  - 1 \right], \nonumber \\
&\equiv& {\nu_p \, L_{\nu_p}  \over 1 - \alpha_c} \left(
C_F^{1-\alpha_c} -1 \right),  \label{Fccompton}
\end{eqnarray}
where the Compton factor $C_F$ has been defined for convenience. 
When  $\alpha_c < 1$, $\dot{m} \gsim 10^{-3} \alpha$ and from 
eq.(\ref{qe+comp}) $\delta \, Q^{+}$ is negligible compared
with $Q^{ie}$.  Therefore $Q^{\rm e+ } \simeq Q^{ie}$. 
Instead of equating $Q^{ie}$ to $P_{\rm Compton}$, and solving for
the temperature, a good approximation to the temperature
can be obtained by rewriting eq. (\ref{alphac})  as a 
quadratic,
\beq
1 + 4\theta_e + 16 \theta_e^2 = \tau_{es}^{-1/\alpha_c},
\eeq
which gives,
\beq
\left( { T_e \over 10^9 } \right) = 0.744 \, 
\left[ \left(4\, \tau_{es}^{-1/\alpha_c} -3 \right)^{1/2} - 1 \right].
\label{tefromalpha}
\eeq
Since $0.5 \le \alpha_c \le 1.0$ for all cases of interest,  we 
can  get an idea for the range of temperatures possible for  a
given $\dot{m}$ by setting $\alpha_c = 0.5$, and $\alpha_c = 1.0$
(lower values of $\alpha_c$ would require very high $\dot{m}$, 
and this is where the advection solutions are no longer valid). 
This gives
\beq
2.15 \times 10^{8} \, \dot{m}^{-1/2} 
\lsim T_e \lsim 
3.12 \times 10^{7} \, \dot{m}^{-1}. \label{terange}
\eeq
For high $\dot{m} \sim 10^{-2}$ systems in this regime, 
eq.(\ref{terange}) indicates that the range of temperatures possible
is confined to $2.15 \times 10^9 \le T_e \le 3.12 \times 10^9$, whether
the systems are $1 M_{\odot}$  or 
$10^9 M_{\odot}$ black holes (cf. Fig. 2). However as $\dot{m}$ 
decreases, while $\alpha_c < 1$,
the temperature range possible increases 
(eg. for $\dot{m} \sim 10^{-2.5}$,  $3.8 \times 10^9 \le
T_e \le 9.8 \times 10^9$),
and a more accurate evaluation of $\alpha_c$ is necessary.
 
We can determine, to a first approximation, the temperature in 
these systems by setting  $\alpha_c \sim 0.75$ in 
eq.(\ref{tefromalpha}).
From this estimate, a more accurate determination
of $\alpha_c$ can be obtained by equating $Q^{ie}$ to 
eq.(\ref{Fccompton}),  and rewriting to give
\beq
1 - \alpha_c = \left. \log \left[ {Q^{ie} \over \nu_p \, L_{\nu_p}} 
\, (1 - \alpha_c) + 1 \right] \right/ \log(C_F).
\eeq
Since logarithms are slowly varying functions, $\alpha_c$ in the
logarithm can be set
to $\sim 0.75$, as before,  to obtain
\beq
1 - \alpha_c \simeq \left. \log \left( {Q^{ie} \over 4 \, \nu_p \, L_{\nu_p}} 
 - 1 \right) \right/ \log(C_F).
\eeq
where $1- \alpha_c$ is the slope of the spectrum on a $\nu \, L_{\nu}$
plot,
\begin{eqnarray}
{Q^{ie} \over \nu_p \, L_{\nu_p}} &=& 3.57 \times 10^2 \, 
\left( {x_M \over 1000} \right)^{-3}
\left( {\alpha \over 0.3} \right)^{-1/2}
\left( {\beta \over 0.5} \right)
\left( {1-\beta \over 0.5} \right)^{-3/2}
\left( {c_1 \over 0.5} \right)^{-1/2}
\nonumber \\
&\times&
\left( {c_3 \over 0.3} \right)^{-1/2}
\left( {r_{\rm min} \over 3} \right)^{3/4}
\left( {T_e \over 10^9} \right)^{-7}
\, g(\theta_e) \, 
m^{1/2} \, \dot{m}^{1/2},
\end{eqnarray}
and
\begin{eqnarray}
C_F &=& 1.46 \times 10^3 \, 
\left( {x_M \over 1000} \right)^{-1}
\left( {\alpha \over 0.3} \right)^{1/2}
\left( {1-\beta \over 0.5} \right)^{-1/2}
\left( {c_1 \over 0.5} \right)^{1/2}
\nonumber \\
&\times&
\left( {c_3 \over 0.3} \right)^{-1/2}
\left( {r_{\rm min} \over 3} \right)^{5/4}
\left( {T_e \over 10^9} \right)^{-1}
\, m^{1/2} \, \dot{m}^{-1/2}.
\end{eqnarray}
Solving for $\alpha_c$ then gives a better 
approximation for the temperature  
from eq.(\ref{tefromalpha}). This  process can be iterated for accurate
determination of both $T_e$ and $\alpha_c$, but we find that fairly
accurate results are obtained without any iterations.
\subsection{Total Luminosity.}
For a given
accretion rate $\dot{M}$, and matter to energy conversion 
of $\eta_{\rm eff} = 0.1$, standard accretion disks predict a total luminosity
of $L_{\rm disk} \simeq \eta_{\rm eff} \, \dot{M} \,  c^2$.
Advection dominated accretion produces a lower luminosity because
most of the viscously dissipated energy is advected inwards with the
flow and deposited into the black hole instead of being
radiated.  The
total luminosity from these disks is equal to the total energy 
being emitted by the various cooling processes, $L_{\rm ADAF} = Q^-$.
However since $Q^{\rm e+} = Q^-$, detailed knowledge of the cooling
processes is not required here, and the total luminosity is
simply $L_{\rm ADAF} = Q^{\rm e+}$. 

Depending on the value of $\dot{m}$, the total heating of the 
electrons is either dominated by $Q^{ie}$ or by $\delta \, Q^{+}$.  The
luminosities in both these regions are determined by 
setting $L_{\rm adv} = {\rm max} (Q^{ie}, \delta \, Q^+)$.  For
$\dot{m} > 10^{-3}\, \alpha^2$ (cf. eq. \ref{qe+comp}), 
$Q^{ie}$ dominates, and the total
luminosity is given by 
\begin{eqnarray}
L_{\rm ADAF} 
&\simeq& 1.2\times 10^{38} \, g(\theta_e) \,
c_1^{-2} \, c_3 \, \beta 
\, r_{\rm min}^{-1}  \, 
\alpha^{-2} 
\, m \, \dot{m}^2 , \nonumber \\
&\simeq& \eta_{\rm eff} 
\dot{M} \, c^2
\left[ \, 0.20 \   
\left( {\dot{m} \over \alpha^2} \right)  \, 
g(\theta_e) \,
\left( { \beta \over 0.5} \right)
\left( { c_1 \over 0.5} \right)^{-2} 
\left( { c_3 \over 0.3} \right)
\left( {  r_{\rm min} \over 3} \right)^{-1}  
\right]
\ \ \mbox{ergs s$^{-1}$},  \nonumber \\
& & 
\label{lumadvection1}
\end{eqnarray}
where $c$ is the speed of light. This also gives the 
luminosity for the case  $\delta = 0$. For $\dot{m} \lsim 10^{-3} \,
\alpha^2$, 
$\delta \, Q^{+}$ dominates the electron heating and the total
luminosity can be written as 
\beq
L_{\rm ADAF} \simeq 
	\eta_{\rm eff} \, \dot{M} \, c^2 \, 
\left[
2.0 \times 10^{-4} \, 
\left( {  \delta  \over 2000^{-1}} \right) 
\left( {  1- \beta  \over 0.5} \right)
\left( {  c_3 \over 0.3} \right)
\left( {  r_{\rm min} \over 3} \right)^{-1} 
\left( {  f \over 1.0} \right)^{-1} 
\right] \ \ \mbox{ergs s$^{-1}$}.
\label{lumadvection2}
\eeq
The factor in the square brackets is the factor by which 
the efficiency is reduced relative to the usual 10\% 
from standard thin accretion disks.  
At
high accretion rates, the luminosity decrease linearly with $\dot{m}$,
but there is no additional 
$\dot{m}$ dependence at low accretion rates since the 
ion--electron transfer rate becomes very inefficient, and the cooling 
processes have to compensate only for a fraction of 
the viscous heating generated.  
Using $L_{\rm ADAF} = Q^{e+}$, and the numerical method in \S 
\ref{numericalmethod},
Fig. 4 shows plots of 
$L_{\rm ADAF}/L_{\rm Edd}$ as a function of $\dot{m}$ for various values
of $\alpha$.  Disks with high values of $\alpha$  are more
sub--Eddington  in their luminosities that disks with low $\alpha$.
At low $\dot{m}$ the luminosities are independent of $\alpha$ (cf.
eq. \ref{lumadvection2}). Although Fig. 4 is calculated for 
$m=10^9$, it can be used for any value of $m$, since  the ratio
$L_{\rm ADAF}/L_{\rm Edd}$ is independent of $m$, and the equilibrium 
temperatures are fairly insensitive to the exact value of $m$. 

\subsection{Determining $\dot{m}_{\rm crit}$.}
In advection flows where $\dot{m} \ll 1$, Narayan \& Yi (1995b)
have shown that  $f\simeq 1$.  However as $\dot{m}$ increases,
the Coulomb interactions
between the ions and electrons become more efficient, and  more 
of the viscously generated energy is transferred from the
ions to the electrons.  This decreases the amount of energy that 
can be advected inwards with the flow, and the value of $f$  therefore
also decreases.  As $\dot{m}$ is increased further, the flow radiates
the generated heat more efficiently, becomes less advection
dominated,  and becomes optically thick.
Narayan \& Yi (1995b) have shown that for $\dot{m}$ greater than 
a critical value, 
$\dot{m}_{\rm crit}$, the energy 
equations (cf. \ref{energyQeq}) 
have no solution, and the advection dominated solution ceases to 
exist. 
Here, we determine  what 
$\dot{m}_{\rm crit}$ is, for a given set of 
parameters $m$, $\alpha$, $\beta$.

To determine the critical accretion rate, 
the energy equation becomes $Q^+(1-f) = Q^- \simeq Q^{ie}$, since 
$Q^{ie} \gg \delta \, Q^+$ in this regime ($\dot{m}\gg 10^{-3} \, \alpha^2$).
Dividing eq.(\ref{qplus}) by 
eq.(\ref{qiemcrit}), and rewriting in terms of $\dot{m}$ gives
\beq
\dot{m} =
7.8 \, {(1-f) \over f} \, {(1-\beta) \over \beta} \, 
\alpha^2 \, c_1^2 \,  { 1 \over g(\theta_e) }.
\eeq
When  $\dot{m} \sim \dot{m}_{\rm crit}$, we expect $f\sim 0.5$, which
requires  about half the generated energy to be radiated away,
which is a reasonable assumption.
Also, for  very high $\dot{m} \sim 10^{-1.7}$, 
eq.(\ref{terange}) shows that
$T_e \sim 1.5\times 10^9$ which gives $g(\theta_e) \sim 7$. 
Setting
$\beta = 0.5$, $c_1 \simeq 0.5$, gives 
\beq
\dot{m}_{\rm crit} \simeq 0.28 \, \alpha^2.
\eeq
The critical accretion rate, in scaled units, is therefore independent
of the mass of the accreting object, but depends quite strongly
on $\alpha$.  This suggests a large value of $\alpha \sim 1$ since 
a low value of $\alpha \sim  0.01$ gives a very small 
$\dot{m}_{\rm crit}$, which is not luminous enough to explain  
even moderate luminosities.  Advection models that have had
success in explaining black hole candidates (eg. 
 Lasota et al. 1996; Narayan et al. 1995) use $\alpha \sim 0.1 
- 0.3$,  and, as suggested by Narayan (1996), could be as high as
 $\sim 1$ to 
explain low luminosity AGNs.
\section{Do Elliptical Galaxies Host Dead Quasars?}
\label{fabiandis}
In this section we use the results above and apply them to a 
specific problem which was first suggested by Fabian \& 
Canizares (1988).  We give a brief introduction to the problem,
derive the results of Fabian \& Canizares (1988), and then, 
as suggested by Fabian \& Rees (1995) show how advection--dominated 
accretion flows resolve this problem. 
\subsection{Outline.}
Most nearby bright elliptical galaxies  
are believed to host `dead' or inactive quasars
(Soltan 1982; Fabian \& Canizares 1988; Fabian \& Rees 1995).
From energetic arguments or from the properties of
broad line emitting regions, the masses of quasars are found to be 
between
$10^8 - 10^{9}$M$_{\odot}$ (Wandel \& Mushotzky 1986).  We should 
therefore
expect black hole masses of this size at the cores of bright
elliptical galaxies, and should be able to detect accretion of the 
ambient gas onto the central black hole.  

From X-ray profiles of elliptical galaxies, we can
determine the density and temperature of the
gas within the central kilo--parsec region.
Since elliptical galaxies are thermally supported, and are
most likely to spherically accrete, we can use
the classical Bondi formula to
obtain what is essentially a lower limit to the luminosity produced by
a black hole of a given mass.  
Fabian \& Canizares (1988) have looked at
six bright nearby ellipticals and, from the observed X-ray luminosity 
of the gas,
have determined upper limits for the black hole masses in these 
galaxies to be $ < 3 \times 10^7$M$_{\odot}$.  This is in conflict
with the expected masses, in these galaxies, 
as determined by Soltan (1982), together with  
the independent estimates of quasar masses as determined
by  Wandel \& Mushotzky (1986).  
One of the conclusions is 
to reject the black hole hypothesis for
quasars, since requiring higher mass black holes, 
would lead to a higher
luminosity in the X-rays, which is not observed.
To reconcile these differences, Fabian \& Rees (1996) have
recently suggested that the massive black holes at the centers
of these galaxies might be undergoing advection dominated 
accretion which would help reconcile these differences.

\subsection{Standard Accretion.}
In this section we show how Fabian \& Canizares (1988) use 
standard Bondi accretion to deduce severe upper limits to the
masses of the black holes at the centers of bright 
elliptical galaxies. 

A lower limit on the accretion 
luminosity is obtained by assuming  that the gas
accretes spherically onto the central black hole 
by Bondi accretion.
Following Fabian \& Canizares (1988), the accretion radius, the 
radius at which the influence of gravity by the central black hole
dominates the dynamics of the gas, is 
$ R_{acc} = \alpha_b \, G \, M / c_s^2 = 
4.32 \ \alpha_b \, M_8 \, T_7^{-1} \ \ \mbox{pc}$,
where $c_s \simeq 10^4 \ T^{1/2}$ cm s$^{-1}$, and $\alpha_b$
is a factor including the ratio of 
specific heats (see Bondi 1952). 
$\alpha_b > 0.5$ and is probably $\sim 1 $. The 
Bondi accretion rate is given by 
$\dot{M} = 
1.86 \times 10^{-4} \ \alpha_b^2 \, P_6 \, T_7^{-5/2} \, M^2_8
                \ \ \mbox{$M_{\odot}$ yr$^{-1}$}$,
where we have written 
$P = n_e \, T = 10^6 P_6$ cm$^{-3}$ K, to keep the 
notation of Fabian \& Canizares (1988). This gives a luminosity
assuming a 10\%  matter to energy conversion, of
\beq
L_b = 1.06 \times 10^{42}  \ \alpha_b^2 \, P_6 \, T_7^{-5/2} \, M^2_8
                \ \ \mbox{ergs s$^{-1}$}.
\label{lb}
\eeq
From eq.(\ref{lb}), if $P_6 = T_7 = 1$, 
black hole masses of $10^8 - 10^{9} M_{\odot}$ should be detectable.

$P_6$ and $T_7$ can be determined by looking at the radial X-ray 
profiles of elliptical galaxies. Canizares et al. (1987) find the 
mean temperatures of the gas to be $\sim 0.5 - 4\times 10^7$ K, and
determine the central number density by calculating the 
volume emissivity of the X-ray gas, $
\epsilon = 4\pi \, n_e(0)^2 \, a_X^2 \, [\ln(2R_X/a_X) -1] \ \
                \mbox{cm$^{-3}$}$,
where $a_X$ is the core radius, and $R_X $ 
is the maximum radial extent of the gas which is chosen to
be $50 \, a_X$.  (This choice is consistent with the radial profiles in
Trinchieri et al. 1986.)   Using the cooling function 
$\Lambda(T) = 10^{-19} \, T^{-1/2}$ ergs cm$^3$ s$^{-1}$,
the central density is 
\begin{eqnarray}
n_e(0)&=&\left( {L_X \over \Lambda(T) \, 4\pi \, a_X^3 \, 3.61} \right)^
{1/2}
                 \ \  \mbox{cm$^{-3}$} \nonumber \\
&=& 4.88 \times 10^{-2} \ \left({L_X \over 10^{41} \, \mbox{ergs s$^{-1}
$}}
                \right)^{1/2}  \,
                \left( {a_X \over 1 \, \mbox{kpc}}\right)^{-3/2} \,
                T_7^{1/4} \ \ \mbox{cm$^{-3}$}
\end{eqnarray}
We assume that the central density $n_e(0)$ evaluated continues on to the
central black hole, i.e. there is no central cavity in these galaxies.
Using $P = n_e \, T$ in eq.(\ref{lb}) the expected
X-ray luminosity from accretion in terms of the total
observed X-ray luminosity from the gas, is given by
\beq
L_b
= 5.17 \times 10^{41} \ \alpha_b^2 \, M_8^2 \, T_7^{1/4} \,
     (a^3 \, T^3_7)^{-1/2} \, \left({L_X \over 10^{41}} \right)^{1/2}
 \ \ \mbox{ergs s$^{-1}$}, \label{lbne}
\eeq
where $a = (a_X / \mbox{1 kpc})$.  This corresponds to an accretion
rate in Eddington units of
\beq
\dot{m} = 4.16 \times 10^{-5} \ \ \alpha_b^2 \, M_8 \ T_7^{1/4} \,
      (a^3 \, T_7^3)^{-1/2} \, \left({L_X \over 10^{41}} \right)^{1/2}.
\label{mdotfabian}
\eeq
Rewriting eq.(\ref{lbne}) to resemble Fabian \& Canizares (1988), and
setting $\alpha_b = 0.5$, we have 
\beq
{L_b \over L_X} = 1.3 \ M_8^2 \left[
        \left({L_X \over 10^{41}} \right) \, a^3 \, T_7^3 \right]^{-1/2}
 \,
                T_7^{1/4} . \label{fabianeq}
\eeq
The quantity $L_b/L_X$ is a measured quantity which is obtained 
by using the X-ray profiles of the elliptical galaxies, and
taking the ratio of the X-ray emission from the 
 central arcsecond region to the total X-ray gas emission from the
whole galaxy. 
Table \ref{galaxytable}  shows the parameters used for 
three of the six galaxies analyzed by
Fabian \& Canizares (1988).  
These galaxies were chosen since they have
good Einstein HRI data (Trinchieri et al. 1986). 
The core X-ray luminosities were estimated
from the surface brightness profiles given in Trinchieri et al. (1986),
taking into account the resolution of the detector.  
The best fits for the core
radius $a_X$ and temperature $T_7$ was also taken from 
Trinchieri et al. (1986).  
Using eq.(\ref{fabianeq}) and the best fit parameters, Table 
\ref{galaxytable} shows the upper limits for black hole masses 
using Bondi accretion. These limits are much too low to be 
consistent with expected masses.

\subsection{Advection--Dominated Accretion.}
We now show, as suggested by Fabian \& Rees (1995),
that advection--dominated accretion resolves this problem.
Using  the scaling laws derived here, we can estimate upper limits
to the black hole masses.  From eq. (\ref{mdotfabian}), we find
that for black hole masses of $\sim 10^{8 - 10} M_{\odot}$, 
$\dot{m} \sim 10^{-3} - 10^{-5}$ and we are in the regime where the 
total luminosity is determined by eq. (\ref{lumadvection1}).  
We also expect $\alpha_c >1$ for these systems, and 
therefore can set $g(\theta_e) \sim 1$ in eq.(\ref{lumadvection1}) 
(cf. eq. \ref{alphg1}).  Setting $c_1 = 0.5$, $c_3 =0.3$, 
$\beta = 0.5$, 
$r_{\rm min} = 3$, eq.(\ref{lumadvection1}) gives
\beq
L_{\rm ADAF}
\simeq   \eta_{\rm eff} \, 0.20 \,  \dot{M} \,
\left( {\dot{m} \over \alpha^2} \right) \, c^2, \ \ \mbox{ergs s$^{-1}$},
\nonumber
\eeq
This is the total luminosity which is emitted over eight orders
in magnitude of frequency.  Assuming that a fraction, $\eta_X$, of 
this energy is radiated into the 0.2 -- 4.0 keV band in the 
X-rays (Trinchieri et al. 1986),  the luminosity in this band
from the advection dominated disk is simply 
$L_{b_{\rm ADAF}} = \eta_X \, L_{\rm ADAF}$.
Multiplying eq.(\ref{lbne}) by $0.20 \, \eta_X \, 
\dot{m}/\alpha^2$ gives 
\beq
{L_{b_{\rm ADAF}} \over L_X} = 4.3 \times 10^{-5} \, \eta_X \, 
        \left({\alpha_b^4 \over \alpha^2} \right) \, M_8^3 \,
        (a^3 T_7^3)^{-1} \, T_7^{-1/2}. \label{madeq}
\eeq
Taking $\alpha_b= 0.5$, as in eq.(\ref{fabianeq}), and $\alpha = 0.3$,
a typical value for advection models, eq.(\ref{madeq}) becomes
\beq
M_8 \simeq 32.2 \  \eta_X^{-1/3}
\left( { \alpha \over 0.3} \right)^{2/3}
\left({L_{b_{\rm ADAF}} \over L_X} \right)^{1/3} \,
	a \,  T_7^{7/6}.
\eeq
The last column in Table \ref{galaxytable} shows upper bounds 
for the masses of the black holes in these galaxies using $\eta_X = 1$,
(a very conservative estimate)
from the advection models. 
The upper limits shown are much higher than those of
Fabian \& Canizares (1988).  A maximum value of $\eta_X$  is obtained 
by arbitrarily setting 
$\alpha_c = 1$.  This gives a flat spectrum on a $\nu L_{\nu}$ plot,
and since the total emission  occurs over eight orders in magnitude of
frequency, and the observations are made only in the 0.2 -- 4.0 keV
band, $\sim 1$ order in magnitude, $\eta_X \lsim 1/8 \simeq  0.13$.
This corresponds to increasing the upper limits in Table 
\ref{galaxytable} by a factor of $\sim 2$. Furthermore 
since $\alpha_c >  1$  in these  systems, 
$\eta_X$ would probably be significantly lower, and 
this would raise the upper limits even more.  Fig. 5 shows the
upper limits of the core X-ray emission, from the galaxies in 
Table \ref{galaxytable}, in the 0.2 -- 4 keV band, 
and shows the spectrum from  
an advection--dominated disk for $m = (0.5, 5, 10, 30) \times 10^8 $,
with the corresponding $\dot{m}$ given by eq.(\ref{mdotfabian}),
and $\alpha=0.3$, $\beta = 0.5$.  Clearly, the value of 
$\eta_X \lsim 0.13$, and easily allows for
 black hole masses $\lsim 10^{10} M_{\odot}$ 
at the centers of bright ellipticals, consistent
with the idea that  bright elliptical galaxies do host dead quasars. 
\section{Discussion \& Conclusion.}
\label{conclude.tex}
The advection models are very robust in that they have
very characteristic spectra:  a $\nu^{1/3}$ slope in the 
radio regime, a sub-mm to X-ray Compton spectrum,  and a 
bremsstrahlung spectrum.  If we assume that a system is 
going through advection--dominated 
 accretion, as in the case of the  elliptical 
galaxies above, we can make predictions of what the spectrum should
look like.  
With $\alpha$, $\beta$ fixed, and $\dot{m}$ given by 
eq.(\ref{mdotfabian}), the only free parameter that can be varied is
the mass. Once this is fixed, the entire spectrum is completely
determined. 
The radio spectrum  in these elliptical galaxies
should follow a $\sim \nu^{1/3}$ 
slope, which extends up to a peak frequency, $\nu_p$.  
Radio observations of these galaxies would therefore 
determine their core masses, and would lead to 
testable predictions for the X-ray fluxes.  
Note that the inclusion of a
thin disk might change the optical and  ultra--violet region of the 
spectrum, but would not affect the strong correlation between the 
radio and X-ray fluxes (eg. Lasota et al. 1996).
Observations in the radio of these elliptical galaxies have been
done (Wrobel 1991).  Although Wrobel (1991) observers weak jets  
at the cores of these elliptical galaxies, 
upper limits to the unresolved  compact core emission has been 
obtained.  
These upper bounds are shown in Fig. 5.  We see that the 
radio bounds  are quite consistent with black hole
masses $m \gsim 10^9$. The masses of NGC 4636, 4649, and 4472, in this
simple description  are
constrained to be less than 10$^9$, $2\times 10^9$, and 
$3\times 10^9 \ M_{\odot}$ respectively.  
This is a remarkable testable
feature of the advection
models:  to explain the entire spectra of these systems using 
few free parameters. 

Interestingly, Slee et al. (1994) have observed radio spectra
in other bright elliptical galaxies, and obtain  
an average radio spectral index of $1/3$.  If this is emission from 
a compact core, it is generally accepted to be from a 
non--thermal source of electrons (eg. Duschl \& Lesch 1994).  
However, if  these low luminosity systems are advection--dominated 
flows,
then the thermal self--absorbed
synchrotron radiation from these models 
naturally give rise to the characteristic 1/3 spectrum produced by
optically thin non--thermal synchrotron emission.

Another interesting application of these models is to explain
low luminosity Active Galactic Nuclei (AGNs, Narayan 1996).  
We again use the strong correlation between the radio and 
X-ray fluxes.
The luminosity of quasars  in the X-rays
are $\gsim 10^{44}$ ergs s$^{-1}$, and their accretion rates
are $> \dot{m}_{\rm crit}$.  However, as the accreting rate 
decreases and falls below $\dot{m}_{\rm crit}$ the accreting 
gas might prefer to follow an advection flow (Narayan \& Yi 1995b).
Since $\dot{m}_{\rm crit}$ is independent of $m$,  all AGNs 
making this transition would have
very hard X-ray spectra with spectral indices $\sim 0.7$,
since $\alpha_c  < 1$.  Since the temperature  of  all systems 
near $\dot{m}_{\rm crit}$ are well determined (cf. Fig. 2 and \S 
\ref{tealphg1}), we can
get a good estimate of $\dot{m}$ using eq.(\ref{alphac}).
This 
could then be combined with the X-ray luminosity to give an estimate
of the mass of the central object. With the mass, accretion rate and
temperature of these systems, the advection--dominated models
can be used to make predictions of the radio spectrum from these
sources. 
Recently Ho (1996) has obtained observations of nearby galaxies which 
show AGN-like  spectral lines, are underluminous, and have
steep X-ray spectra.  Observations in the radio of these galaxies
would not only serve as a test for the advection models, but would
also independently determine the masses of the central objects.

We have shown that the general properties of optically
thin  advection dominated flows can be easily understood through
simple scaling laws.  The spectra that these models produce can
be reproduced fairly well from a basic knowledge of the three
electron cooling processes. For high $\dot{m}$, the Compton power
is the dominant source of cooling which gives a very hard X-ray 
spectrum.  In the opposite limit, for low $\dot{m}$, synchrotron
cooling dominates the cooling, and most of the energy is emitted
in the radio.  The bremsstrahlung power is negligible, but depending
on the amount of Compton power, can dominate the X-ray emission.

These results have been applied to determine, as suggested by
Fabian \& Canizares (1988), and more recently by 
Fabian \& Rees (1996),  whether dead quasars are at the centers of
elliptical galaxies.  We have found that if these are 
advection--dominated  
systems, which is most likely due to the low accretion rates, then
the upper limits are much higher $M \lsim 5 \times 10^{9} M_{\odot}$
than that determined by Fabian \& Canizares (1988) 
$M \lsim 3 \times 10^{7} M_{\odot}$. Therefore we are allowed
to have black hole masses of $M \lsim 10^{10} M_{\odot}$ at the
centers of bright ellipticals as required by independent 
arguments.

The advection models are constantly tested by observations.  
Since there are few 
free parameters in the model, and the predicted spectrum ranges over
all observable frequencies,  failure to comply with any observation 
would pose serious problems.  All the observations on putative 
$1 M_{\odot}$  to $10^{9} M_{\odot}$   advection dominated 
black hole systems have
so far agreed quite well with predictions.  

\noindent{\em Acknowledgments:} The author thanks Ramesh Narayan
for many useful discussions and comments, and the referee A. C. Fabian
for helpful suggestions. This work was supported
by NSF grant AST 9423209.
\newpage
\begin{appendix}
\section{Analytic Approximation to $q^{ie}$.} \label{appendixqie}
The energy transfer rate from the ions to electrons via Coulomb 
collisions is given by Stepney \& Guilbert (1983)
\begin{eqnarray}
\lefteqn{q^{ie} = 5.61\times 10^{-32} \ {n_e^2 \, (T_i - T_e) \over
		K_2(1/\theta_e)\, K_2(1/\theta_i)} } \nonumber \\
&\times& \left[ {2(\theta_e +\theta_i)^2 +1 \over (\theta_e + \theta_i)} \, 
K_1\left({\theta_e + \theta_i \over \theta_e \, \theta_i} \right ) + 
2 K_0\left( {\theta_e + \theta_i \over \theta_e \, \theta_i} \right )
\right] \ \  \mbox{ergs cm$^{-3}$ s$^{-1}$.}  \label{qie2}
\end{eqnarray}
The following identities hold for the temperature range of interest,
\beq
10^9 < T_i < 10^{12}, \mbox{\hspace{0.5cm}} 
	10^{-4} < \theta_i < 10^{-1}, 
\mbox{\hspace{0.5cm}} 10 < \theta_i^{-1} <  10^4,
\eeq
and 
\beq
10^9 < T_e < 10^{10}, \mbox{\hspace{0.5cm}}
        0.17 < \theta_e < 1.7,
\mbox{\hspace{0.5cm}} .6 < \theta_e^{-1} <  6.
\eeq
The arguments of the modified Bessel functions $K_0$ and
$K_1$  are  large for these values of $\theta_e$ and $\theta_i$ 
which enable the use of the 
approximation (Abramowitz \& Stegun 1964, 9.7.2) 
\beq
K_n(x) \simeq \sqrt{{\pi \over 2 \, x}} \, e^{-x} \, 
	\left( 1 + {4n^2 - 1 \over 8 \, x}  + \ldots  \right).
\eeq
Since $\theta_i \ll 1$, terms of order $O(\theta_i/\theta_e)$ can be 
neglected. This gives
\begin{eqnarray}
K_0\left({\theta_e + \theta_i \over \theta_e \, \theta_i} \right) 
& \simeq &  \sqrt{{ \pi \over 2}} \, \left( {\theta_e \, \theta_i 
	\over \theta_e + \theta_i} \right)^{1/2} \, 
		e^{-1/\theta_i}  \, e^{-1/\theta_e} \,  \\
K_1\left({\theta_e + \theta_i \over \theta_e \, \theta_i} \right)
& \simeq &  \sqrt{{ \pi \over 2}} \, \left( {\theta_e \, \theta_i
\over \theta_e + \theta_i }\right)^{1/2} \,
  e^{-1/\theta_i}  \, e^{-1/\theta_e} \,  \\
K_2\left({1 \over \theta_i} \right) &\simeq &  \sqrt{{ \pi \over 2}}\,
\theta_i^{1/2} \, e^{-1/\theta_i}.
\end{eqnarray}
Eq.(\ref{qie2}) then becomes
\begin{eqnarray}
\lefteqn{q^{ie} \simeq 5.61\times 10^{-32} \ {n_e\, n_i \, (T_i - T_e) \over
                K_2(1/\theta_e)} }\nonumber \\
&\times& \left( {\theta_e \, \theta_i \over \theta_i 
( \theta_e + \theta_i)} \right)^{1/2} \, \left[
{2(\theta_e +\theta_i)^2 +1 + 2(\theta_e + \theta_i) \over 
(\theta_e + \theta_i)} \right] \, e^{-1/\theta_e} \ \  
\mbox{ergs cm$^{-3}$ s$^{-1}$,} 
\end{eqnarray}
which simplifies to 
\beq
q^{ie} \simeq 5.61\times 10^{-32} \ n_e\, n_i \, (T_i - T_e) \, 
g(\theta_e) \ \ \mbox{ergs cm$^{-3}$ s$^{-1}$}.
\eeq
where 
\beq
g(\theta_e) \equiv {1 \over
                K_2(1/\theta_e)} \, \left(2 + 2\theta_e + { 1 \over
                \theta_e} \right) \, e^{-1/\theta_e}.
\eeq
Values of $g(\theta_e)$ are given in Table 1.
\section{\bf Determining $x_M$} \label{xmappendix} 
From eq.(\ref{xmeq})  we have
\beq
\exp\left(1.8899 \, x_M^{1/3}\right) = 2.49 \times 10^{-10} \ 
	{4 \pi \, n_e \, R \over B} \, {1 \over \theta_e^3 \, 
		K_2(1/\theta_e) } \, \left( {1 \over x_M^{7/6}} 
			+ {0.40 \over x_M^{17/12}} + {0.5316 \over
				x_M^{5/3}} \right). \label{xmeq2}
\eeq
Since most systems of interest are highly self-absorbed,  
$x_M$ will be large, and therefore fairly independent of $r$.
\footnote{Numerical calculations have shown that $x_M \sim r^{1/15}$, 
for $r \lsim 10^3$.} 
In this case, we can set $r = 3$ in eq.(\ref{xmeq2}), and 
  neglect the last two terms in 
the parentheses (this can be checked for self consistency).
Substituting for $n_e$, $R$, and $B$ from eqs.(\ref{alleq}) in 
eq.(\ref{xmeq2}), and taking logarithms on both sides, gives
\begin{eqnarray}
y + 1.852 \ln y &\simeq& 
10.36
+  0.26 \ln \left( m \, \dot{m} \right) 
- 0.26\ln \left[ \theta_e^3 \, K_2\left(1/\theta_e \right) \right]
\nonumber \\
&-& 0.26\ln \left[
\left({\alpha \over 0.3}\right)
\left({c_1 \over 0.5}\right)
\left({c_3 \over 0.3}\right)
\left({1 - \beta \over 0.5}\right)\right].
\label{yeq}
\end{eqnarray}
where $$y=x_M^{1/3}.$$ 
This equation can be solved numerically, and 
Table \ref{tablet3k} shows the values of 
$\theta_e^3 \, K_2\left(1/\theta_e \right)$ for the temperature range
of interest. 
Fig. 2 shows plots of $x_M$ as  a function 
of $\dot{m}$ for different values of black hole mass $m$, where the
value of $x_M$ is determined after solving for the 
equilibrium temperature in the flows (cf. \S \ref{numericalmethod}).
Since $x_M$ is weakly dependent on $m$, $\alpha$, $\beta$, but 
depends mainly on $\dot{m}$, we have a useful formula for the
dependence of $x_M$ on $\dot{m}$ 
\beq
\log x_M = 3.6 + {1 \over 4} \log \dot{m},
\eeq
which can be used 
for different values of $m$, $\alpha$ and $\beta$ to a 
good approximation. 
\section{Formulae for $\delta = 0$.}
\label{tedetappendix}
In this appendix we give formulae for $\delta = 0$.
In  \S \ref{tealphg1}  we obtained an equation for the 
temperature for $\alpha_c >1$ where we neglected $Q^{ie}$ compared
with $\delta \, Q^{+}$.  If $\delta = 0 $ or $\dot{m} \ge 10^{-4}$ 
the temperature has to be determined by setting 
\begin{eqnarray}
Q^{ie} = Q^- &\simeq& \left( 0.71 + {1 \over \alpha_c -1} \right) \,
        \nu_p \, L_{\nu_p} \nonumber \\
&\simeq& A_c \, \nu_p \, L_{\nu_p},
\end{eqnarray}
where the first term is due to synchrotron cooling and the second is 
due to Compton cooling.  Using eq. (\ref{qiemcrit}) and rewriting, gives 
\beq
{T_e^7 \over g(\theta_e)} \simeq {1.2\times 10^{74} \over A_c}
\, x_M^{-3} \,
\alpha^{-1/2} \, \beta \, (1-\beta)^{-3/2} \, c_1^{-1/2} \, c_3^{-1/2}
\,
m^{1/2} \, \dot{m}^{1/2} \,
r_{\rm min}^{3/4}.
\eeq
To simplify further, $g(\theta_e)$ can be approximated  to
\beq
g(\theta_e) \simeq 1.91\times 10^{11} \, T_e^{-1.1464},
\eeq
which is valid for $10^9$ K $ \le T_e \le 3\times 10^{10}$ K, and has
a maximum error of 20\% at $T_e \sim 10^9$.
\footnote{The error made in this approximation is reduced when 
taking the $\sim 1/7^{th}$ power to determine $T_e$.}
Using this approximation and canonical values of the constants gives
\begin{eqnarray}
T_e &\simeq&
{2.7 \times 10^{9} \over A_c^{3/25} }
\left( {x_M \over 1000} \right)^{-2/5}
\left( {\alpha \over 0.3} \right)^{-3/50}
\left( {\beta \over 0.5} \right)^{3/25}
\left( {1-\beta \over 0.5} \right)^{-1/5}
\left( {c_1 \over 0.5} \right)^{-3/50} \nonumber \\
&\times&
\left( {c_3 \over 0.3} \right)^{-3/50}
\left( {r_{\rm min} \over 3} \right)^{1/10} \nonumber
 m^{3/50} \, \dot{m}^{3/50} \ \ \mbox{K}, \label{appendixte1}
\end{eqnarray}
where $0.96 \le A_c^{3/25} \le 1.3$, and 
we have approximated the exponents to the nearest fraction.
\end{appendix}
\newpage
\begin{table}
\caption[Modified Bessel function]{$\theta_e$ and $K_2(1/\theta_e)$.}
\label{tablet3k}
\begin{center}
\begin{tabular}{lccl}\hline \label{k2table} 
& & & \\
\rb{$T_9 $} & \rb{$\theta_e$} & \rb{$g(\theta_e)$} &
\rb{$ \theta_e^{3} \, K_2(1/\theta_e)$} \\[-.5ex] \hline \hline
1.00 & 0.1686 & 12.003 & 8.783e--06 \\ 
1.50 & 0.2530 & 6.7292 & 2.982e--04 \\ 
2.00 & 0.3373 & 4.5134 & 2.472e--03 \\ 
2.50 & 0.4216 & 3.3386 & 1.092e--02 \\ 
3.00 & 0.5059 & 2.6261 & 3.408e--02 \\ 
3.50 & 0.5902 & 2.1540 & 8.550e--02 \\ 
4.00 & 0.6746 & 1.8209 & 1.849e--01 \\ 
4.50 & 0.7589 & 1.5746 & 3.593e--01 \\ 
5.00 & 0.8432 & 1.3859 & 6.438e--01 \\ 
5.50 & 0.9275 & 1.2369 & 1.083e+00 \\ 
6.00 & 1.0118 & 1.1166 & 1.731e+00 \\ 
6.50 & 1.0961 & 1.0175 & 2.654e+00 \\ 
7.00 & 1.1805 & 0.9345 & 3.930e+00 \\ 
7.50 & 1.2648 & 0.8640 & 5.650e+00 \\ 
8.00 & 1.3491 & 0.8035 & 7.922e+00 \\ 
8.50 & 1.4334 & 0.7509 & 1.086e+01 \\ 
9.00 & 1.5177 & 0.7048 & 1.462e+01 \\ 
9.50 & 1.6021 & 0.6641 & 1.933e+01 \\ 
10.00 & 1.6864 & 0.6278 & 2.519e+01 
\end{tabular}
\end{center}
\end{table}
\newpage
\begin{table}
\caption[Galaxies of Fabian \& Canizares (1988)]
{Galaxies analyzed from Fabian \& Canizares (1988). Distances are
taken from Trinchieri et al. (1986).}
\begin{center}
\begin{tabular}{lcccccccc}\hline \label{galaxytable}
& & & & & & & \\
\rb{Galaxy} & \rb{Distance}& \rb{$M_B$} & \rb{$a_X$} & \rb{$T_7$} & \rb{$\log(L_X)$} & \rb{$L_b/L_X$} & \rb{10$^8 \, M_{\odot}$} &
\rb{10$^8 \, M_{\odot}$} \\
\rb{NGC} &\rb{Mpc}& & \rb{kpc} & & & & \rb{(FC)} & \rb{(Advection)} \\[-.5ex]
\hline \hline
4472 & 20 & -22.8 & 0.48 & 1.4 & 41.71 & 0.025 & 0.14  & 6.7 \\
4649 & 20 & -22.2 & 0.96 & 1.4 & 41.40 & 0.047 &  0.29 & 24.1 \\
4636 & 16.4 & -21.6 & 1.18 & 1.2 & 41.64 &  0.030 & 0.27 & 14.6 \\
\end{tabular}
\end{center}
\end{table}
\clearpage
{
\footnotesize
\StartRef
\noindent {\large \bf References} \\
\Ref Abramowitz, M., \& Stegun, A. I., 1965, Handbook of Mathematical 
Functions (National Bureau of Standards, Washington, D. C.) \\
\Ref Abramowicz, M., Chen, X., Kato, S., Lasota, J. P, \& Regev, O., 1995,
ApJ, 438, L37 \\
\Ref Abramowicz, M., Czerny, B., Lasota, J. P, \& Szuszkiewicz, E., 1988,
ApJ, 332, 646 \\
\Ref Bondi, H., 1952, MNRAS, 112, 195-204 \\
\Ref Canizares, C. R., Fabbiano, G., \& Trinchieri, G., 1987, ApJ, 312, 
503-513 \\
\Ref Dermer, C. D., Liang, E. P., \& Canfield, E., 1991, ApJ, 369, 410 \\
\Ref Duschl, W., \& Lesch, H., 1994, A\& A, 286, 431-436 \\
\Ref Esin, A., in preparation \\
\Ref Fabian, A. C., \& Canizares, C. R., 1988, Nature, 333, 829-831 \\
\Ref Fabian, A. C., \& Rees, M. J., 1995, MNRAS, 277, L55-L58 \\
\Ref Frank, J., King, A., \& Raine, D., 1992, Accretion Power in 
Astrophysics (Cambridge: Cambridge Univ. Press) \\
\Ref Haswell, A. C., Robinson, L. E., Horne, K., Stiening, F. R., 
Abbott, M. C. T., 1993, ApJ, 411, 802-812 \\
\Ref Ho, C. L., 1996, in The Physics of LINERS in View of Recent 
Observations, eds. Eracleous, M., Koratkar, P. A., Ho, C. L.,  \&
Leitherer, C., (San Francisco: ASP) \\
\Ref Lasota, J. P., Abramowicz, M. A., Chen, X., Krolik, J.,
                Narayan, R., \& Yi, I. 1996, ApJ, in press \\
\Ref Mahadevan, R., Narayan, R., \& Yi., I., 1996, ApJ, 465, 327-337 \\
\Ref Matsumoto, R., Kato, S., \& Fukue, J., 1985, in Proc. Symposium on 
Theoretical Aspects of Structure, Activity, and Evolution of 
Galaxies: III, ed. S. Aoki, M. Iye, and Y. Yoshii, (Tokyo: Tokyo 
Astron. Obs.), 102 \\
\Ref Narayan, R., 1996, ApJ, 462, 136 \\
\Ref Narayan, R., McClintock, J. E., \& Yi, I., 1996, ApJ, 457, 821-833 \\
\Ref Narayan, R., \& Yi, I., 1994, ApJ, 428, L13 \\
\Ref Narayan, R., \& Yi, I., 1995a, ApJ, 444, 231 \\
\Ref Narayan, R., \& Yi, I., 1995b, ApJ, 452, 710-735 \\
\Ref Narayan, R., Yi, I., \& Mahadevan, R., 1995, Nature, 374, 623-625 \\
\Ref Narayan, R., Yi, I., \& Mahadevan, R., 1996, A\&A, in press \\
\Ref Rees, M. J., Begelman, M. C., Blandford, R. D., \& Phinney, E. S., 
1982, Nature, 295, 17 \\
\Ref Rybicki, G., \& Lightman, A., 1979, Radiative Processes in 
Astrophysics (New York: John Wiley \& Sons, Inc.) \\
\Ref Shapiro, S. L., Lightman, A. P., \& Eardley, D. M. 1976, ApJ, 204, 187 \\
\Ref Slee, O. B., Sadler, E. M., Reynolds, J. E., \& Ekers, 1994, MNRAS, 269
, 928-946 \\
\Ref Stepney, S., \& Guilbert, P. W., 1983, MNRAS, 204, 1269 \\
\Ref Soltan, A., 1982, MNRAS, 200, 115-222 \\
\Ref Trinchieri, G., Fabbiano, G., Canizares, C. R., 1986, ApJ, 310, 
637-659 \\
\Ref van der Marel, R., 1995a, MNRAS, 274, 884-898
\Ref van der Marel, 1995b, in New Light on Galaxy Evolution, 
(ed. Bender, R., \& Davies, R. L.), IAU Symp. No. 171, Kluwer, 
Heidelberg. \\
\Ref Wandel, A., \& Mushotzky, R. F., 1986, ApJ, 306, L61-L66 \\
\Ref Wrobel, M. J., 1991, AJ, 101, 127-147\\

}
\newpage
\noindent{\bf Figure Captions} \\
Figure 1: The spectrum produced by an advection--dominated disk with 
$\alpha = 0.3$, $\beta = 0.5$, $m = 5 \times 10^9$, and $ \dot{m} 
= (3, 6, 12, 24) \times 10^{-4}$.  The plots are calculated numerically
by the method described in \S \ref{numericalmethod}.
The three labels correspond to the three cooling processes: 
synchrotron cooling (S), Compton cooling (C), and bremsstrahlung 
cooling (B).  $\nu_p$ and $\nu_{\rm min}$ correspond to the 
radio frequencies from the region $ 3 \le r \le 10^3$.

\noindent Figure 2:
The equilibrium temperatures as a function of 
$\dot{m}$, for different values of $m$,  
and the corresponding values of $x_M$. 
For low $\dot{m}$, 
$\delta \, Q^+$ dominates the heating of the electrons.

\noindent Figure 3:
Plot of $1-\alpha_c$ as a function of $\dot{m}$ for the 
corresponding plots in Fig. 2.  

\noindent Figure 4:
Plot of $L_{\rm ADAF}/L_{\rm Edd}$ as a function of $\dot{m}$ for
different values of $\alpha$. The plot can be used for any value
of $m$ (see text).

\noindent Figure 5:
Spectra for $m = (0.5, 5, 10, 30) \times 10^8$, and their
corresponding $\dot{m}$ given by eq.(\ref{mdotfabian}).
 The bar and arrow represent the 0.2 -- 4 keV upper
bounds for the x-ray core emission from the bright elliptical 
galaxies given in Table \ref{galaxytable}.  The upper bounds
in the radio are the unresolved compact core fluxes (Wrobel 1991).
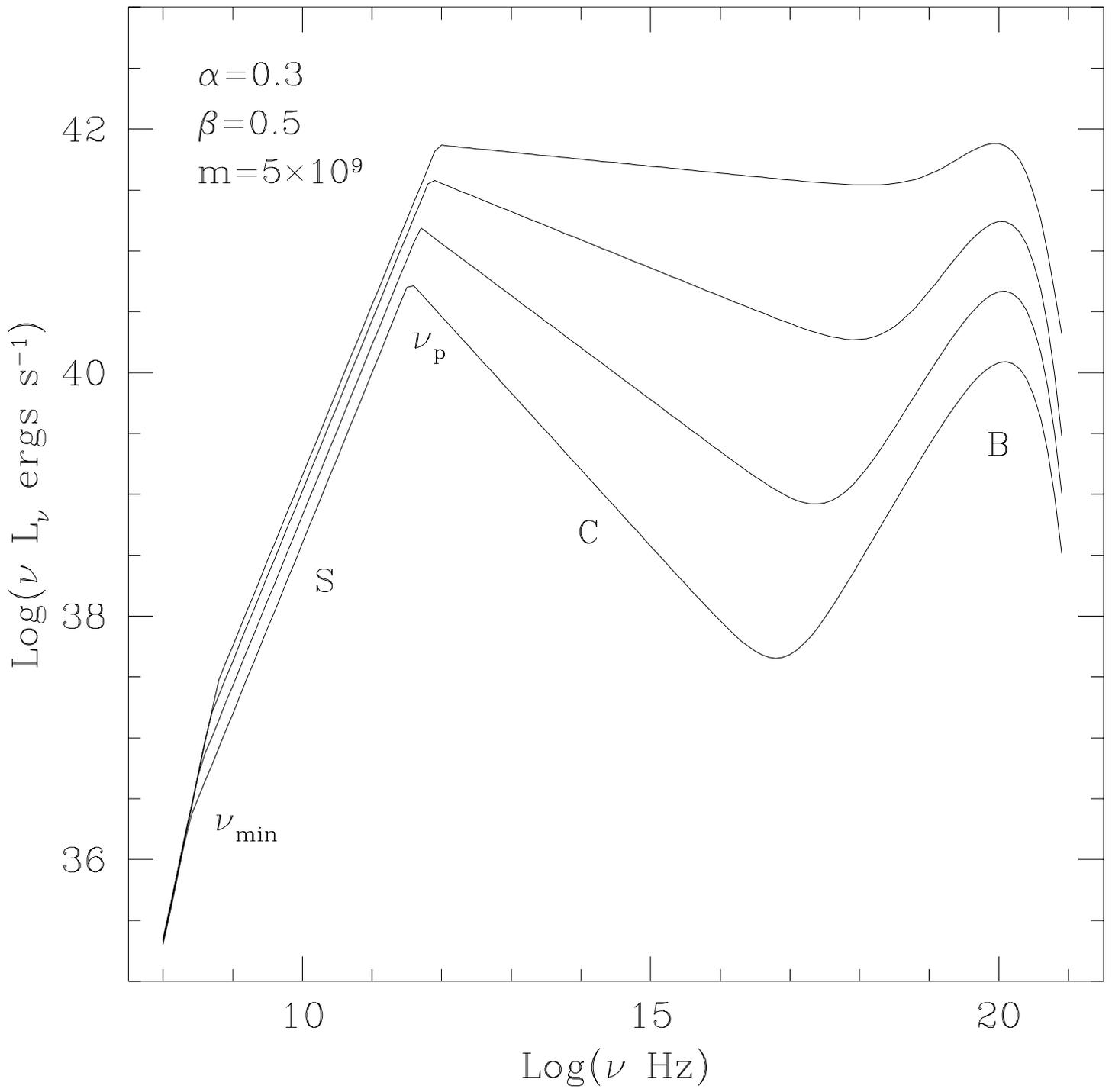
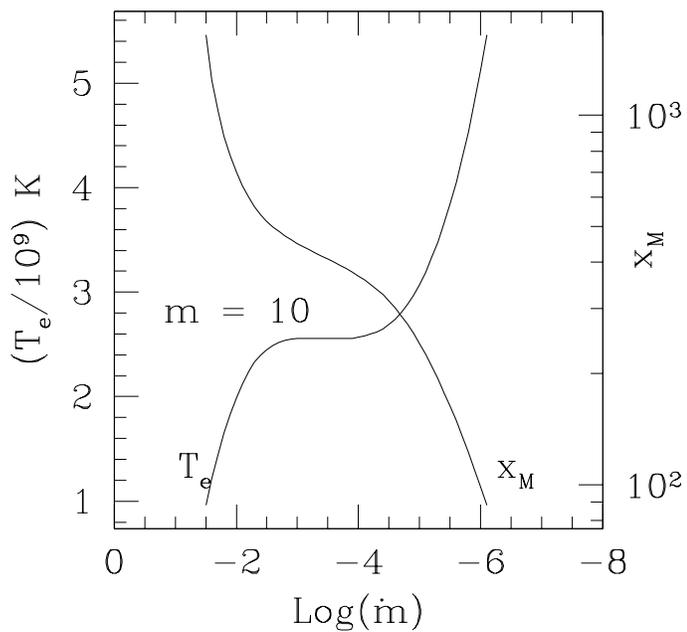
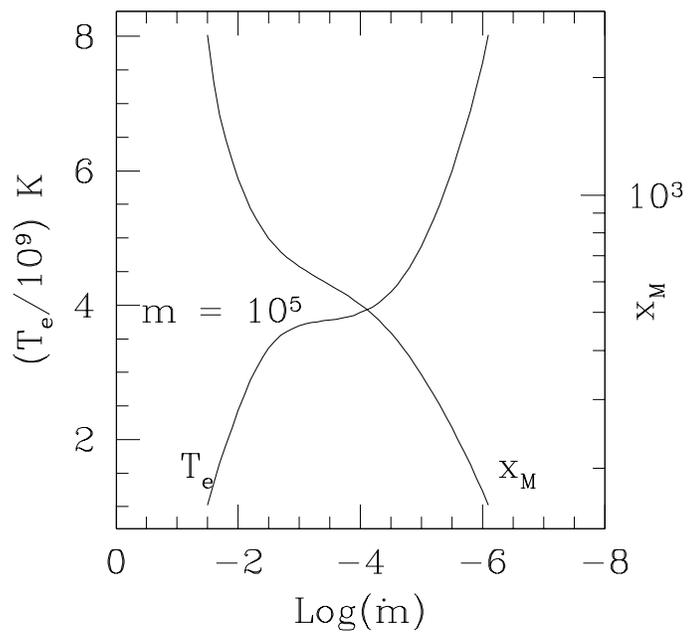
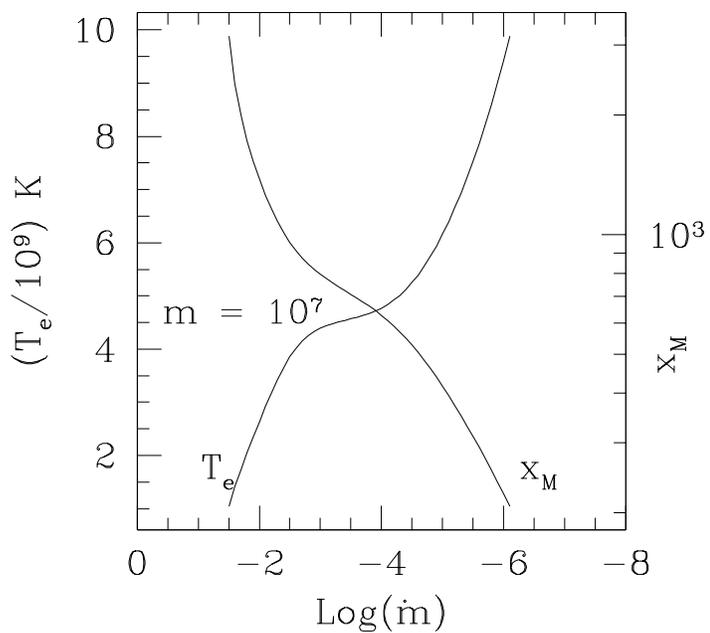
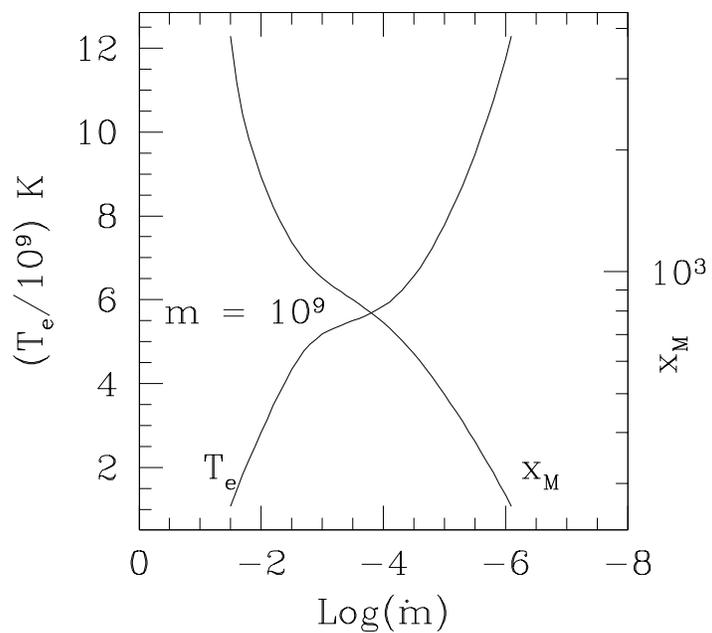
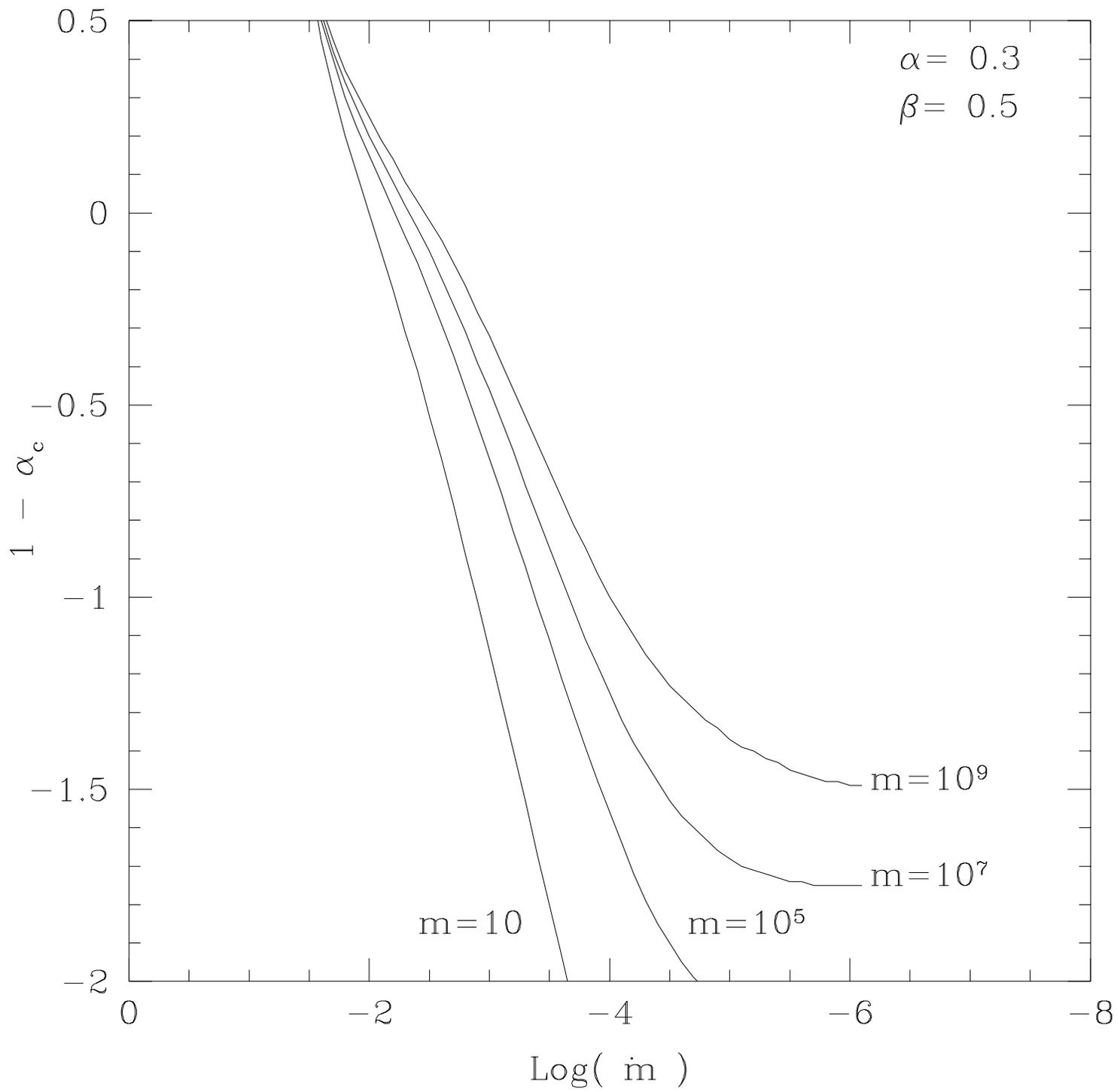
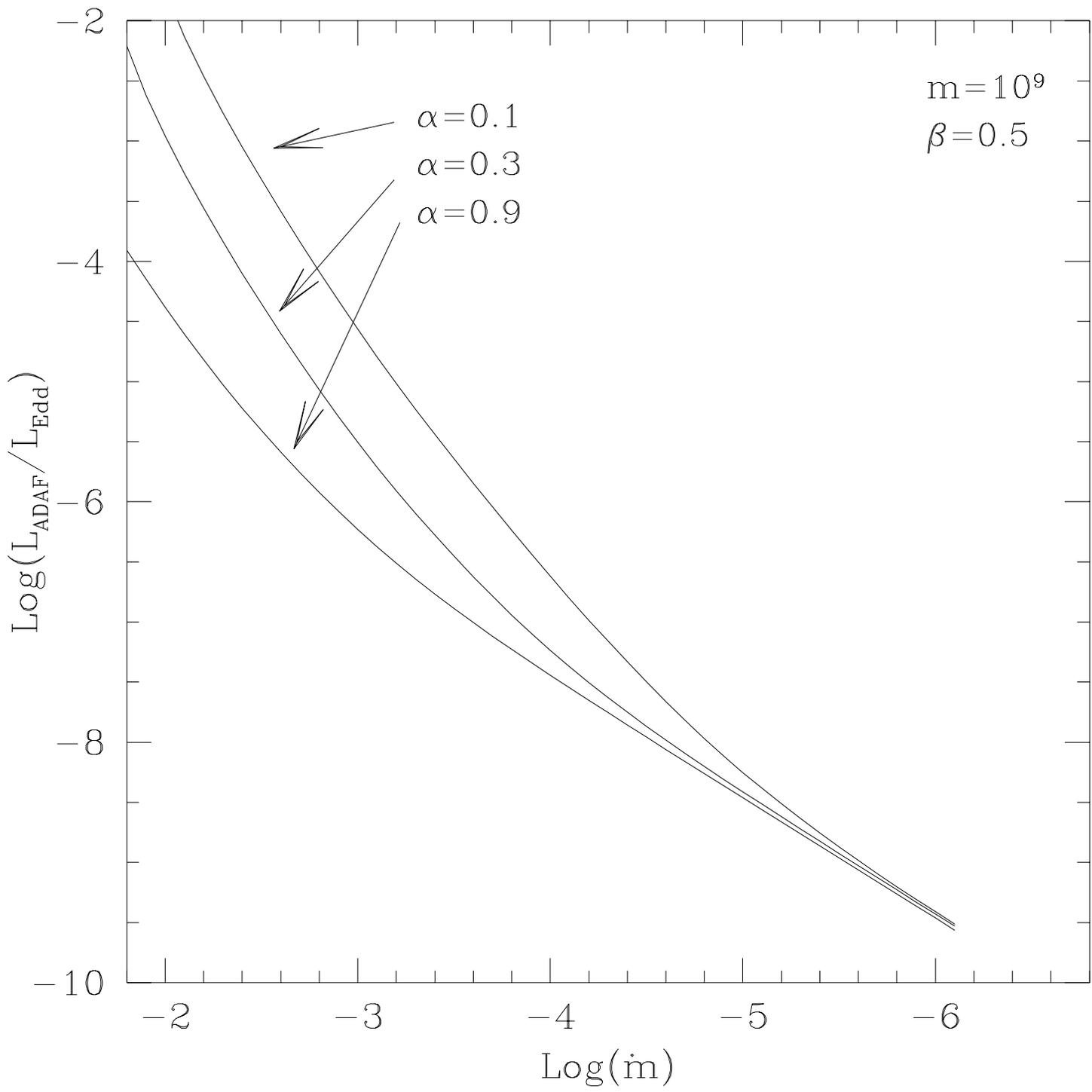
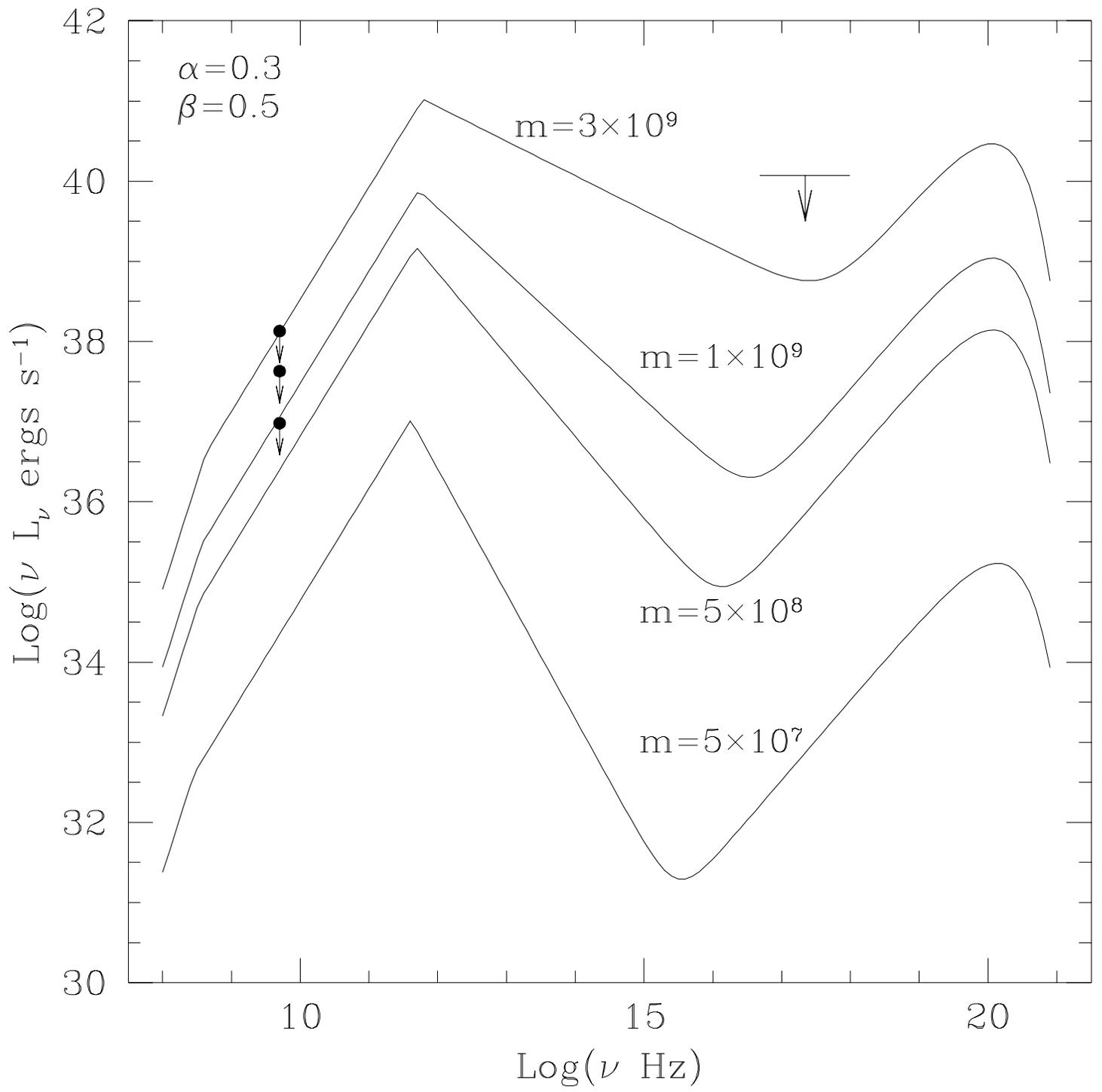
\end{document}